\documentclass[
 reprint,
 amsmath,amssymb,
 aps,
 prstab,
 nofootinbib
]{revtex4-1}

\usepackage{graphicx}
\usepackage{dcolumn}
\usepackage{bm}
\usepackage{siunitx}
\usepackage{xcolor}
\usepackage{ulem}
\usepackage{subcaption}
\usepackage{hyperref}

\begin{document}

\title{Nonlinear equilibria and emittance growth in plasma wakefield accelerators with ion motion}

\author{C. Hansel}
\email{claire.hansel@ucla.edu}
\author{W. An}
\author{W. Mori}
\author{J. B. Rosenzweig}
\affiliation{
 Department of Physics and Astronomy, University of California, Los Angeles, California 90095 
}

\date{\today}

\begin{abstract}

The plasma wakefield accelerator may accelerate particles to high energy in a future linear collider with unprecedented acceleration gradients, exceeding the \si{GeV/m} range. Beams for this application would have extremely high brightness and, subject to the intense plasma ion-derived focusing, they would achieve densities high enough to induce the plasma ions to collapse into the beam volume. This non-uniform ion density gives rise to strong nonlinear focusing which may lead to deleterious beam emittance growth. The effects of ion collapse and their mitigation has been investigated recently through particle-in-cell simulations, which show that by dynamically matching the beam to the focusing of the collapsed ion distribution, one may avoid serious emittance growth. We extend this work by exploring the near-equilibrium state of the beam-ion system reached after the ions have collapsed, a condition yielding the emittance growth mitigation observed. We show through PIC simulations  \cite{an} and analytical theory that in this case a dual electron beam-ion Bennett-type equilibrium distribution is approached. Here, the beam and ion distributions share nearly the same shape, which generates nonlinear transverse electromagnetic fields. We exploit a Bennett-type model to study beam phase space dynamics and emittance growth over time scales much longer than permitted by PIC simulations through use of a 2D symplectic tracking code with Monte Carlo scattering based on Moliere's theory of small angle multiple scattering. We find that while phase space diffusion due to parametric excitations of the beam size due to plasma non-uniformity is negligible, scattering from collapsed ions gives rise to manageable emittance growth in the case of a linear collider. The implications of these results on experiments planned at FACET-II are examined.

\end{abstract}

\maketitle

\section{Introduction}

Due to radio-frequency cavity breakdown \cite{cahill} and related phenomena, current conventional particle linear accelerators are limited to a maximum practical acceleration gradient not notably in excess of 100 MeV/m. To diminish the size of high energy accelerators in applications exemplified by the linear collider (LC), a number of advanced acceleration techniques have been proposed and are in various stages of development.  One such technique is plasma wakefield acceleration, which employs waves in a plasma - a medium chosen to naturally evade breakdown issues - excited by an intense drive beam, to accelerate a trailing beam. Plasma wakefield accelerators (PWFAs) have already demonstrated acceleration gradients in excess of 50 \si{GeV/m} \cite{nature50gevm}, orders of magnitude above what is achievable with conventional accelerators. Further scenarios have been identified that extend this gradient reach to beyond a TeV/m \cite{tevpermeter,Manwani2020}. To explore the suitability of the PWFA for LC applications, proposals \cite{afterburner1, afterburner2} have been put forward that analyze the use of an ``afterburner" at the end of a conventional LC injector, which have the goal of doubling the beam energy available for high energy physics exploration. There is considerable worldwide research interest in development of the PWFA for this high impact application, with major new facilities now being commissioned to explore the physics issues supporting PWFA development such as FACET-II \cite{facet2} and FLASHForward \cite{flashforward}.

Current experimental and theoretical research on PWFAs has focused on the nonlinear ``blowout" regime due to its favorable properties for acceleration and focusing. In the blowout regime the electron beam density is much greater than the ambient plasma electron density, and the collective fields of the beam eject (or ``blow out") the plasma electrons from the region near the beam path, and form a bubble of negligible electron density \cite{jamieblowout}. An  electromagnetic wave is trapped inside of this bubble that provides acceleration in uniform phase fronts, as in standard relativistic electron accelerators. Further, in this scenario the plasma ions left behind, if undisturbed, provide a uniform charged column that yields strong, linear (emittance preserving) focusing. In this way, one may achieve high quality, low energy spread acceleration without emittance growth due to geometric aberrations. However, in the proposed PWFA afterburner case, the low emittance, high current LC beams acting under the influence of strong ion focusing will move significantly during beam passage. The plasma ions, previously treated as stationary with constant density $n_0$, can then rearrange dramatically, collapsing into a tight column with much enhanced, non-uniform density \cite{jamieionmotioncalc}. This in turn causes the beam to focus further under the influence of large, nonlinear forces, potentially inducing notable emittance growth. 

The challenge of controlling such emittance growth has been known for over a decade, but recently a robust solution has been found through an analysis based on particle-in-cell (PIC) simulations \cite{an}. This work demonstrates that that by matching the initial beam phase space to the eventual focusing arising from the collapsed ion distribution, a quasi-equilibrium scenario develops, in which emittance growth is mitigated to a nearly ignorable level. This quasi-equilibrium has been investigated further by Benedetti, \textit{et al.}, \cite{benedetti, benedetti2} in a transient regime, where its nature is not clear, as the result of the analysis is a complicated analytical description of the beam-plasma state. Here, we extend these previous explorations by examining the fundamental characteristics of this matched, nonlinear equilibrium, in the limit where it is established (collisionlessly, through nonlinear phase mixing) in a near steady state over the majority of the beam. The analytical form of this equilibrium is available through a Maxwell-Vlasov description, and is found to be a Bennett-type distribution where the ion and electron distributions follow nearly the same radial profile. The establishment of this equilibrium is described theoretically and verified through PIC simulations. 

Due to the computationally intensive nature of these simulations, the evolution of the beam's phase space can only be modeled on a short time scale of a few betatron oscillations. As a consequence, PIC simulations cannot model effects that take place over longer timescales such as diffusion induced by the nonlinear focusing or scattering due to the high plasma ion density. We have developed a tracking code that exploits the analytically known focusing forces in equilibrium to investigate the electron beam behavior over timescales much longer than those that can be feasibly simulated with PIC codes. In these tracking simulations, we are concerned with two effects: diffusion in phase space due to parametric excitations and Coulomb scattering due to the very dense (over 10 times atmospheric density) plasma ions. We use this code to estimate the observability of phase space dilution and concomitant emittance growth in a relevant LC case, and for the parameters of the E314 Ion Motion experiment currently under development at the SLAC wakefield research facility FACET-II. We discuss aspects of this experiment in light of the physics understanding developed from simulation and theory. 

With these issues in mind, we organize this paper as follows. In Sec.~(\ref{sec:bennett}) we derive the Maxwell-Vlasov equilibrium equations that yield the Bennett-type distributions, and give expressions for the the Bennett-type profile parameters expected in terms of beam and plasma initial conditions. In Sec.~(\ref{sec:collapsesim}) we display and discuss particle-in-cell simulations that illustrate the collapse to equilibrium and validate the description of the equilibrium as having Bennett-type form. In Sec.~(\ref{sec:eqsim}) we describe the tracking code that permits evaluation of long-time phase space dynamics in the beam. In Sec.~(\ref{sec:stability}) we discuss the possibility, based on the use of this tracking code, of observing diffusion due to chaos and parametric resonances driven by plasma density fluctuations. Finally in Sec.~(\ref{sec:conclusion}) we discuss conclusions, implications for present experiments, and future concerns related to a PWFA-based LC. Additionally, Appendix \ref{sec:multiplescattering} contains a derivation of the multiple scattering angular distribution as well as a description of how the tracking code was validated.

\section{Bennett-type Equilibria} \label{sec:bennett}

In this section, we develop the analysis of the Maxwell-Vlasov equilibrium that arises due to this two species -- electron beam and collapsed ion population -- interaction. It is in many ways an extension and generalization of the magnetically self-focused beam scenario originally analyzed by Bennett in 1934.

\subsection{Equilibrium Conditions}

We begin with a brief description of the Maxwell-Vlasov analysis leading to the emergence of a Bennett-type equilibrium. Let $n_e$ and $n_i$ be the number densities of beam electrons and plasma ions respectively which are assumed to be cylindrically symmetric. The beam is taken to travel with approximately constant velocity $\beta c \hat{z}$, \textit{i.e.} predomininantly in the $z$-direction. The plasma ions are assumed to be in a thermal equilibrium without a net directed flow in $z$. The total charge density is thus given by $\rho = -e n_e + Z e n_i$ and the total current density by $\vec{J} = -e \beta c n_e \hat{z}$. The electromagnetic fields are then obtained in excellent approximation by using Gauss's law and Amp\`ere's law, as is customary in ultra-relativistic beam analyses. From these fields, the forces on a beam electron and on a plasma ion are obtained from the Lorentz force relation, and the resulting expressions are simplified using the assumptions $\dot{r} \ll c$ and $\dot{z} \simeq \beta c$ for beam electrons and $\dot{r} \ll c$ and $\dot{z} \ll c$ for plasma ions. The total force on a beam electron is thus

\begin{equation} \label{eq:eforce}
\vec{F}_e(\vec{r}) = \frac{-e^2}{\epsilon_0 r} \int_0^r \left(Z n_i(r) - \frac{n_e(r)}{\gamma^2} \right) r dr \hat{r}.
\end{equation}

\noindent By assuming $Z n_i \gg n_e / \gamma^2$, which holds for GeV beams, this expression can be simplified to

\begin{equation} \label{eq:eforceapprox}
\vec{F}_e(\vec{r}) = \frac{-Ze^2}{\epsilon_0 r} \int_0^r n_i(r) r dr \hat{r}.
\end{equation}

\noindent The force on a plasma ion is, on the other hand, 

\begin{equation} \label{eq:iforce}
\vec{F}_i(\vec{r}) = \frac{-Ze^2}{\epsilon_0 r}\int_0^r \left( n_e(r) - Z n_i(r) \right) r dr \hat{r}.
\end{equation}

The Hamiltonian yielding such forces on a beam electron or plasma ion can now be written:

\begin{equation} \label{eq:hamiltonian}
\mathcal{H}_{\alpha} = \frac{p_x^2}{2 \gamma_\alpha m_\alpha} + \frac{p_y^2}{2 \gamma_\alpha m_\alpha} - \int F_{\alpha, r}(r) dr,
\end{equation}

\noindent where the subscript $\alpha$ indicates either $e$ for beam electron or $i$ for plasma ion. Note that $\gamma_e = \gamma$ and $\gamma_i \simeq 1$ since the ions are assumed to move non-relativistically. To obtain the equilibrium distribution function $f_\alpha (\vec{q}, \vec{p})$, rather than solving the Maxwell-Vlasov equation directly as done in Ref.~\cite{jamiebennett}, we exploit the fact that this thermal (globally uniform temperature) form can be generally written as $f_\alpha(\vec{q}, \vec{p}) \propto e^{-\frac{\mathcal{H_\alpha}(\vec{q}, \vec{p})}{\tau_\alpha}}$ where $\tau_\alpha$ is the transverse temperature of the species. Thus

\begin{equation}
f_\alpha(x, y, p_x, p_y) = C_\alpha e^{-\frac{p_x^2 + p_y^2}{2 \gamma_\alpha m_\alpha \tau_\alpha} + \frac{1}{\tau_\alpha} \int F_{\alpha, r}(r) dr}.
\end{equation}

 \noindent We now explicitly place $f_{\alpha}$ in in separable form by performing a coordinate transformation to yield

\begin{equation}
\begin{split}
f_\alpha(r, \theta, p_x, p_y) &= \lambda_{\alpha} f_{r, \alpha}(r) f_{\theta, \alpha}(\theta) f_{p_x, \alpha}(p_x) f_{p_y, \alpha} (p_y) \\
f_{r, \alpha}(r) &= C_\alpha e^{\frac{1}{\tau_\alpha} \int F_{\alpha, r}(r) dr} \\
f_{\theta, \alpha}(\theta) &= (2\pi)^{-1} \\
f_{p_x, \alpha}(p_x) &= \frac{1}{\sqrt{2\pi}\sigma_{p_x, \alpha}}e^{-\frac{p_x^2}{2\sigma_{p_x, \alpha}^2}} \\
f_{p_y, \alpha}(p_y) &= \frac{1}{\sqrt{2\pi}\sigma_{p_y, \alpha}}e^{-\frac{p_y^2}{2\sigma_{p_y, \alpha}^2}}
\end{split}
\end{equation}

\noindent where $\lambda_{\alpha}$ is the number of particles of species $\alpha$ per unit longitudinal length, $\sigma_{p_x, \alpha} = \sigma_{p_y, \alpha} = \sqrt{\gamma_\alpha m_\alpha \tau_\alpha}$, and $f_\alpha$ is normalized such that

\begin{equation} \label{eq:fnormalization}
\begin{split}
& \int_0^{\infty}f_{r, \alpha}(r)r dr = \int_0^{2\pi}f_{\theta, \alpha}(\theta) d\theta = \\ & = \int_{-\infty}^{\infty} f_{p_x, \alpha}(p_x) d p_x = \int_{-\infty}^{\infty} f_{p_y, \alpha}(p_y) d p_y = 1.
\end{split}
\end{equation}

\noindent The equilibrium condition for each species is obtained from the fact that the radial component off the equilibrium distribution must be proportional to the number density of that species. Mathematically,

\begin{equation} \label{eq:eqcond}
n_\alpha(r) = C_\alpha e^{\frac{1}{\tau_\alpha} \int F_{\alpha, r}(r) dr},
\end{equation}

\noindent where $C_\alpha$ has been redefined to absorb the constant of proportionality. 

\subsection{Bennett Profile} \label{subsec:bennett}

Now we examine the Bennett-type equilibrium obtained from this analysis. Combining Eq.~(\ref{eq:eforceapprox}) with Eq.~(\ref{eq:eqcond}) and taking $\alpha = e$ gives an integral equation for the beam electron density $n_e$. Similarly combining Eq.~(\ref{eq:iforce}) with Eq.~(\ref{eq:eqcond}) and taking $\alpha = i$ gives an integral equation for the plasma ion density $n_i$. It is straightforward to verify that the densities

\begin{equation} \label{eq:ebennett}
n_e = \frac{\rho_e}{(1+(\frac{r}{a})^2)^2}
\end{equation}

\noindent and

\begin{equation} \label{eq:ibennett}
n_i = \frac{\rho_i}{(1+(\frac{r}{a})^2)^2}
\end{equation}

\noindent solve this system of integral equations, provided the temperatures are given by

\begin{equation} \label{eq:bennetttaueapprox}
\tau_e = \frac{Z e^2 a^2 \rho_i}{8 \epsilon_0}
\end{equation}

\noindent and

\begin{equation}
\tau_i = \frac{Z e^2 a^2 (\rho_e - Z\rho_i)}{8 \epsilon_0}.
\end{equation}

\noindent Substituting Eq.~(\ref{eq:ebennett}) into Eq.~(\ref{eq:eforce}) yields the force on a beam electron,

\begin{equation} \label{eq:eforcebennett}
\vec{F}_e(r) = -\frac{Z e^2 \rho_i}{2 \epsilon_0} \frac{\vec{r}}{1+\left(\frac{r}{a} \right)^2}.
\end{equation}

\noindent Similarly, substituting Eq.~(\ref{eq:ibennett}) into Eq.~(\ref{eq:iforce}) yields the force on a plasma ion

\begin{equation} \label{eq:iforcebennett}
\vec{F}_i(r) = -\frac{Ze^2(\rho_e - Z \rho_i)}{2 \epsilon_0} \frac{\vec{r}}{1+\left(\frac{r}{a} \right)^2}.
\end{equation}

\noindent Substituting these forces into into Eq.~(\ref{eq:hamiltonian}) yields the Hamiltonian, which for beam electrons is

\begin{equation} \label{eq:ehamiltonianbennett}
\begin{split}
\mathcal{H}_e &= \frac{p_x^2}{2 \gamma m_e} + \frac{p_y^2}{2 \gamma m_e} + \\ &+ \frac{Z e^2 a^2 \rho_i}{4 \epsilon_0} \ln \left( 1 + \left(\frac{r}{a} \right)^2 \right)
\end{split}
\end{equation}

\noindent and for plasma ions is

\begin{equation} \label{eq:ihamiltonianbennett}
\begin{split}
\mathcal{H}_e &= \frac{p_x^2}{2m_i} + \frac{p_y^2}{2m_i} + \\ &+ \frac{Ze^2 a^2(\rho_e - Z \rho_i)}{4 \epsilon_0} \ln \left( 1 + \left(\frac{r}{a} \right)^2 \right).
\end{split}
\end{equation}

\noindent The beam electron and ion distribution functions are given in this analysis by

\begin{equation} \label{eq:bennettf}
\begin{split}
& f_\alpha(r, \theta, p_x, p_y) = \lambda_{\alpha} \times  \left(\frac{2a^2}{\left(a^2 + r^2 \right)^2} \right) \times \left( \frac{1}{2 \pi} \right) \times \\ & \times \left( \frac{1}{\sqrt{2\pi}\sigma_{p_x, \alpha}}e^{-\frac{p_x^2}{2\sigma_{p_x, \alpha}^2}} \right) \times \left( \frac{1}{\sqrt{2\pi}\sigma_{p_y, \alpha}}e^{-\frac{p_y^2}{2\sigma_{p_y, \alpha}^2}} \right)
\end{split}
\end{equation}

\noindent where

\begin{equation} \label{eq:lambdatorho}
\lambda_{\alpha} = \pi a^2 \rho_{\alpha}
\end{equation}

\noindent and

\begin{equation} \label{eq:fsigma}
\sigma_{p_x, \alpha} = \sigma_{p_y, \alpha} = \sqrt{\gamma_\alpha m_\alpha \tau_\alpha}.
\end{equation}

\subsection{Modified Bennett Profile} \label{subsec:modifiedbennett}

As discussed in the following section, PIC simulations show that while the beam density is accurately described by a Bennett profile, the ion density is not so satisfactorily described. Indeed, while the center of the ion column is relatively well approximated by a Bennett profile, the density profile shown in Fig.~\ref{fig:collapse} exhibits ``wings" at a distance of a few Bennett radii off axis as well as a constant background ion density ``pedestal" for $r \gg a$. While accounting for the wing shape is outside the scope of the following analysis, it is important to explore an analytical description of the resulting changes to the beam profile due to the constant background ion density that extends outside the collapsed region.

We begin our analysis by defining the {\it modified} ion Bennett profile as the original ion Bennett profile Eq.~(\ref{eq:ibennett}) plus a uniform background ion density $n_0$:

\begin{equation} \label{eq:modifiedibennett}
n_i = \frac{\rho_i}{(1+(\frac{r}{a})^2)^2} + n_0.
\end{equation}

\noindent To derive the modified electron Bennett profile we start with the equilibrium Eq.~(\ref{eq:eqcond}) with $\alpha = e$. Substituting in the force Eq.~(\ref{eq:eforceapprox}), we obtain an equation for $n_e$. We make the assumption that the temperature $\tau_e$ in this equation is independent of the background ion density $n_0$ -- the ions are approximately in a global thermal equilibrium after the collapse process. 

With this assumption, and since the modified electron Bennett profile should reduce to the unmodified electron Bennett profile Eq.~(\ref{eq:ebennett}) for $n_0 \rightarrow 0$, $\tau_e$ can be determined. The modified electron Bennett profile is thus given by

\begin{equation} \label{eq:modifiedebennett}
n_e = \frac{\rho_e e^{-\frac{r^2}{2 \sigma_{\textup{tail}}^2}}}{(1+(\frac{r}{a})^2)^2}
\end{equation}

\noindent where

\begin{equation}
\sigma_{\textup{tail}} = \frac{a}{2}\sqrt{\frac{\rho_i}{n_0}}
\end{equation}

\noindent and

\begin{equation}
\tau_e = \frac{Z e^2 a^2 \rho_i}{8 \epsilon_0}.
\end{equation}

The modified Bennett profiles, in contrast to the original Bennett form, are only an approximate equilibrium because they do not self-consistently solve for the ion equilibrium. However, as the errors in the modified description are quantitatively scaled to the undisturbed ion density, which is small compared to the final peak ion density. Thus the error arising from this approach should be negligible.  To further provide justification of this assertion, we point to the fact the the form of the modified ion Bennett profile Eq.~(\ref{eq:modifiedibennett}) agrees well with the results of the PIC simulations (except for the distribution wings), as shown in Fig.~\ref{fig:collapse}.

The existence of the ion density pedestal in equilibrium results in a major conceptual benefit: while Eq.~(\ref{eq:ebennett}) possesses a divergent second moment $\langle x^2 \rangle$,  Eq.~(\ref{eq:modifiedebennett}) yields a bounded second moment, and the rms emittance of the electron beam in the modified Bennett case is no longer infinite. This is also a helpful property for the tracking code discussed in Sec.~(\ref{sec:eqsim}). If the rms emittance diverges, the emittance computed by the tracking code yields an unphysical dependence on the number of particles tracked. Despite this important conceptual difference, the radial dependence of modified Bennett distribution is nearly indistinguishable in practice from the Bennett profile, as illustrated in Fig.~\ref{fig:collapse}. This is owed to the fact that only the  tails of the distribution are affected by the ion pedestal. In this regard, the feature that the force is linear in $r$ far off axis rather than decaying as $r^{-1}$ makes phase space diffusion less of a concern. Such diffusion processes are discussed at length in Sec.~(\ref{sec:stability}).

\subsection{Determination of Bennett Radius}

The goal of this section is to derive expressions for the three parameters $a$, $\rho_e$, and $\rho_i$ of the beam electron and plasma ion Bennett distributions Eqs.~(\ref{eq:ebennett},\ref{eq:ibennett}) derived from the initial conditions of the system. Three equations are required to determine these three parameters.

The first expression comes from the conservation of the number of beam electrons per unit longitudinal length $\lambda_e$ during the process of collapse. This, along with Eq.~(\ref{eq:lambdatorho}) yields 

\begin{equation} \label{eq:first}
\lambda_{e, \textup{initial}} = \pi a^2 \rho_e.
\end{equation}

Similarly, the second equation comes from the fact that the number of plasma ions per unit longitudinal length should be constant during the process of collapse. However since the plasma ions are initially uniformly distributed, the initial number of ions per unit longitudinal length is, before considering which ones may be involved in the interaction, infinite. To resolve this issue, we include as relevant to the analysis only plasma ions with an initial distance from the axis less than some value $r_i$. The initial number of ions per unit length is thus $\pi r_i^2 n_0$ and from this along with Eq.~(\ref{eq:lambdatorho}), we obtain 

\begin{equation} \label{eq:second}
n_0 r_i^2 = \rho_i a^2.
\end{equation}

The value of $r_i$ is estimated through the following considerations. The equation of motion for a plasma ion can be obtained utilizing the force given by Eq.~(\ref{eq:iforcebennett}):

\begin{equation}
\vec{r}'' + \frac{Z e^2 (\rho_e - Z \rho_i)}{2 \epsilon_0 m_i \beta^2 c^2} \frac{\vec{r}}{1 + \left(\frac{r}{a} \right) ^2} = 0,
\end{equation}

where derivatives are respect to $z = \beta c t$. For $r \ll a$ the motion is simple harmonic with angular wave-number

\begin{equation} \label{eq:ki}
k_i^2 = \frac{Z e^2 (\rho_e - Z \rho_i)}{2 \epsilon_0 m_i \beta^2 c^2}.
\end{equation}

We use this frequency to quantify the time needed for nonlinear phase mixing that yields the equilibrium. For this purpose, we take as distance for ion equilibration to be approximately two linear ion oscillation periods. Further, We take $r_i$ to be as the maximum initial radius of an ion that can fall to a final radius of one Bennett radius in this distance. Assuming negligible angular momentum and initial radial velocity, an expression for $r'(z)$ of a particle starting at $r = r_i$ can be derived from Eq.~(\ref{eq:ihamiltonianbennett}) employing conservation of energy. Defining $\tilde{r} \equiv r / a$, $\tilde{r}_i \equiv r_i / a$, and $\tilde{z} \equiv k_i z$, this expression becomes

\begin{equation}
\frac{d\tilde{r}}{d\tilde{z}} = \sqrt{\ln \left( \frac{1 + \tilde{r}_i^2}{1 + \tilde{r}^2} \right)}.
\end{equation}

\noindent Since $\tilde{r}$ goes from $\tilde{r} = \tilde{r}_i$ to $\tilde{r} = 1$ in a distance $\tilde{z} = 4\pi$,

\begin{equation}
4 \pi = \int_{1}^{\tilde{r}_i} \frac{1}{\frac{d\tilde{r}}{d\tilde{z}}} d\tilde{r} = \int_{1}^{\tilde{r}_i} \left( \ln \left( \frac{1 + \tilde{r}_i^2}{1 + \tilde{r}^2} \right) \right) ^ {-\frac{1}{2}} d\tilde{r}.
\end{equation}

\noindent Solving this expression numerically we obtain $\tilde{r}_i = 10.24$.

The third and final equation is obtained by assuming there is no dilution of the distribution density in the center of the beam's transverse phase space during the process of collapse to equilibrium. This condition will be enforced in the case of the stronger condition that there is negligible transverse emittance growth during the establishment of the equilibrium. While the lack of distribution density dilution is clearly not always a valid assumption, it has been shown in Ref.~\cite{an} that it is possible to match the initial beam size so that emittance growth due to collapse is nearly ignorable. With negligible beam-core phase space dilution, the value of the distribution function $f(\vec{r}, \vec{p})$ at $\vec{r} = \vec{p} = 0$ is approximately conserved. If we take the initial beam spatial distribution to be described by a cylindrically symmetric bi-gaussian function, the initial value of $f(\vec{0}, \vec{0})$ is

\begin{equation} \label{eq:fzzinitial}
f_{e, \textup{initial}}(\vec{0}, \vec{0}) = \frac{\lambda_{e, \textup{initial}}}{4 \pi^2 \sigma_x^2 \sigma_{p_x}^2} = \frac{m_e^2 c^2 \lambda_{e, \textup{initial}}}{4 \pi^2 \epsilon_n^2}
\end{equation}

\noindent where we have used the fact that the normalized rms emittance of a cylindrically symmetric bi-gaussian distribution is $\epsilon_n = \sigma_x \sigma_{p_x} / m_e c$. From Eq.~(\ref{eq:bennettf}), the value of $f(\vec{0}, \vec{0})$ after collapse is

\begin{equation} \label{eq:fzzfinal}
f_{e, final}(\vec{0}, \vec{0}) = \frac{\lambda_{e, final}}{2 \pi^2 a^2 \sigma_{p_x, e} \sigma_{p_y, e}} = \frac{4 \epsilon_0 \rho_e}{\pi \gamma m_e Z e^2 a^2 \rho_i}
\end{equation}

\noindent where we have used Eqs.~(\ref{eq:bennetttaueapprox}), (\ref{eq:lambdatorho}), and (\ref{eq:fsigma}). To obtain the third and final equation we set Eqs.~(\ref{eq:fzzinitial}) and (\ref{eq:fzzfinal}) equal to eachother to get

\begin{equation} \label{eq:third}
\frac{\rho_e}{a^2 \rho_i} = \frac{r_e m_e^4 c^4 Z \gamma  \lambda_{e, \textup{initial}}}{4 \epsilon_n^2}.
\end{equation}

\noindent Combining Eqs.~(\ref{eq:first}), (\ref{eq:second}), and (\ref{eq:third}) we obtain

\begin{equation} \label{eq:rhoeexp}
\rho_e = \frac{\lambda_e \tilde{r}_i}{2 \epsilon_n} \sqrt{\frac{r_e \gamma Z n_0}{\pi}}
\end{equation}

\begin{equation} \label{eq:rhoiexp}
\rho_i = n_0 \tilde{r}_i^2
\end{equation}

\begin{equation} \label{eq:bennettexp}
a = \left( \frac{4 \epsilon_n^2}{\pi r_e \gamma Z n_0 \tilde{r}_i^2} \right)^{\frac{1}{4}}
\end{equation}

\noindent where

\begin{equation}
\tilde{r}_i \approx 10.24
\end{equation}

This is a physically intuitive result, as it is similar to the expression for the square of rms beam size $\sigma^2$ in the case of linear focusing; this spot area is proportional to the the rms emittance ${\epsilon_n/\gamma}$ multiplied by the inverse of the linear-focusing betatron wave-number $k_\beta$. In the case of ion focusing, $k_\beta=\sqrt{2\pi r_e Z \rho_i/\gamma}$, and the scaling $a\propto\sigma\propto \sqrt{\epsilon_n/\gamma k_\beta}$ is manifested in Eq.~(\ref{eq:bennettexp}).

\section{Ion Collapse Simulation: Demonstration of Near-Equilibrium Distributions} \label{sec:collapsesim}

\begin{figure*}
\centering
\begin{subfigure}[h]{\columnwidth}
\centering
\includegraphics[width=\columnwidth]{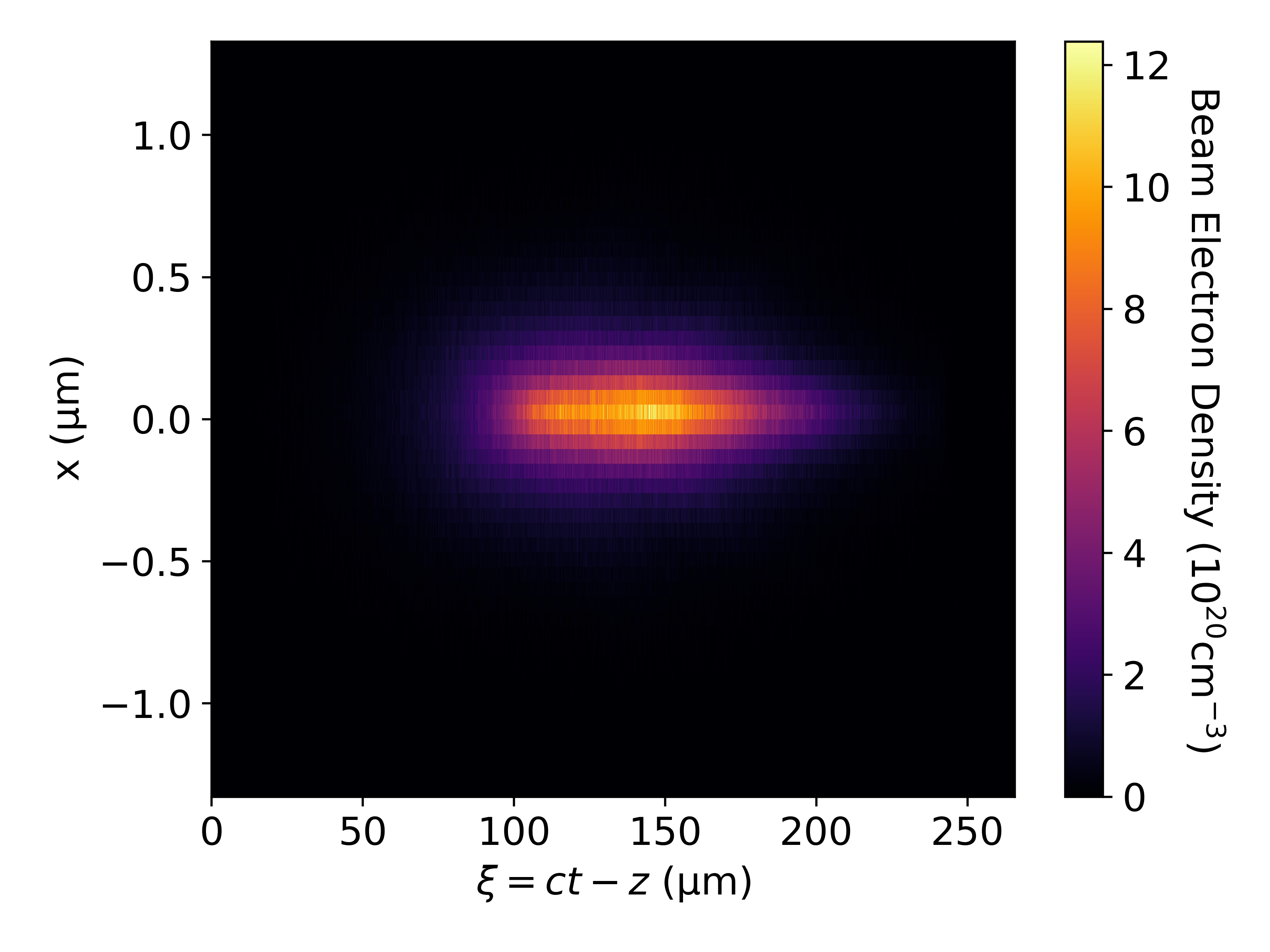}
\caption{}
\end{subfigure}
\begin{subfigure}[h]{\columnwidth}
\centering
\includegraphics[width=\columnwidth]{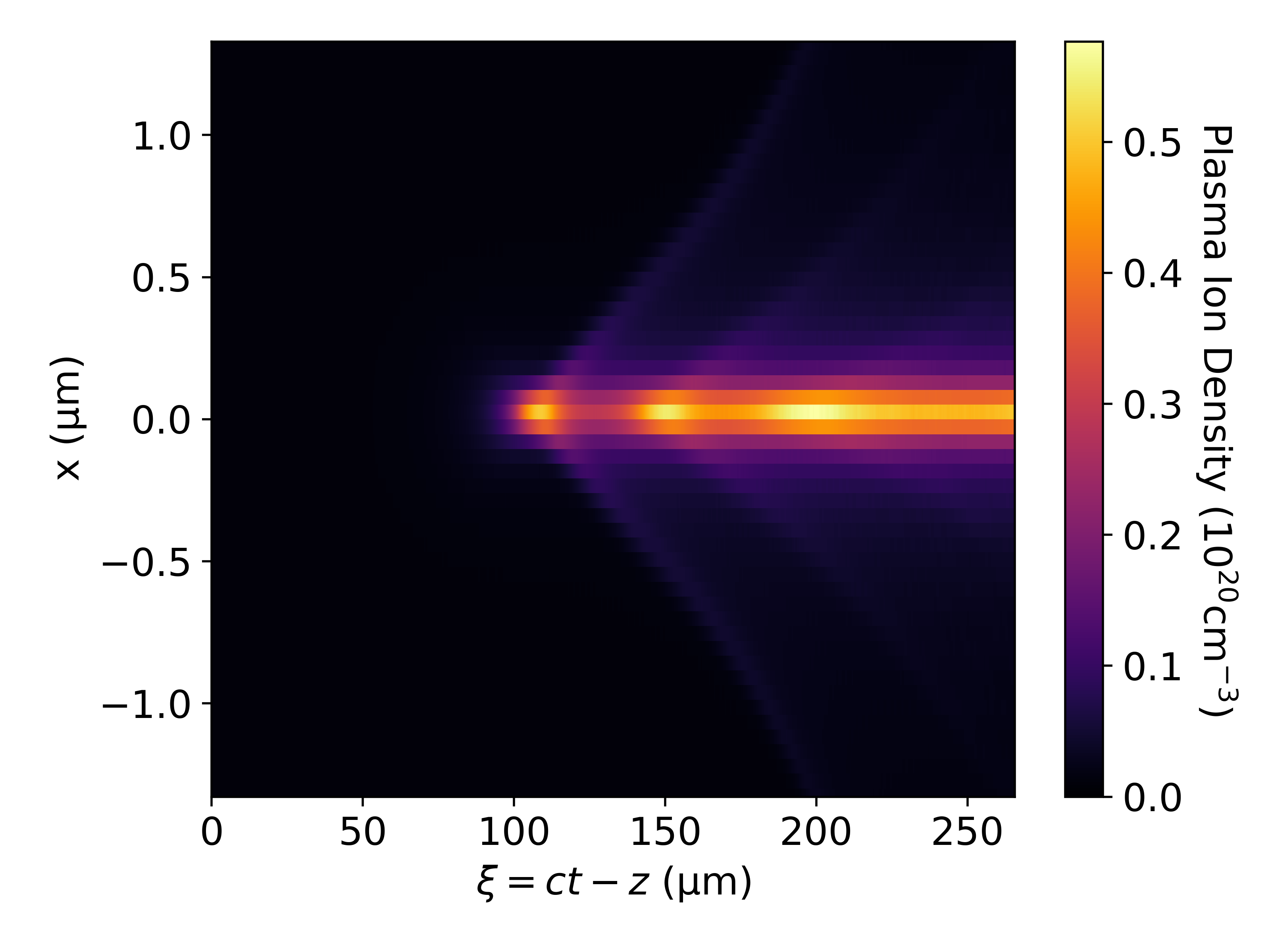}
\caption{}
\end{subfigure}
\caption{Densities of beam electrons (a) and plasma ions (b) in the 2D slice $y = 0$ at the end of the ion collapse simulation. Simulation performed using QuickPIC with the parameters shown in Table \ref{tab:colsimparams}.}
\label{fig:heatmap}
\end{figure*}

\begin{figure*}
\centering
\begin{subfigure}[h]{\columnwidth}
\centering
\includegraphics[width=\columnwidth]{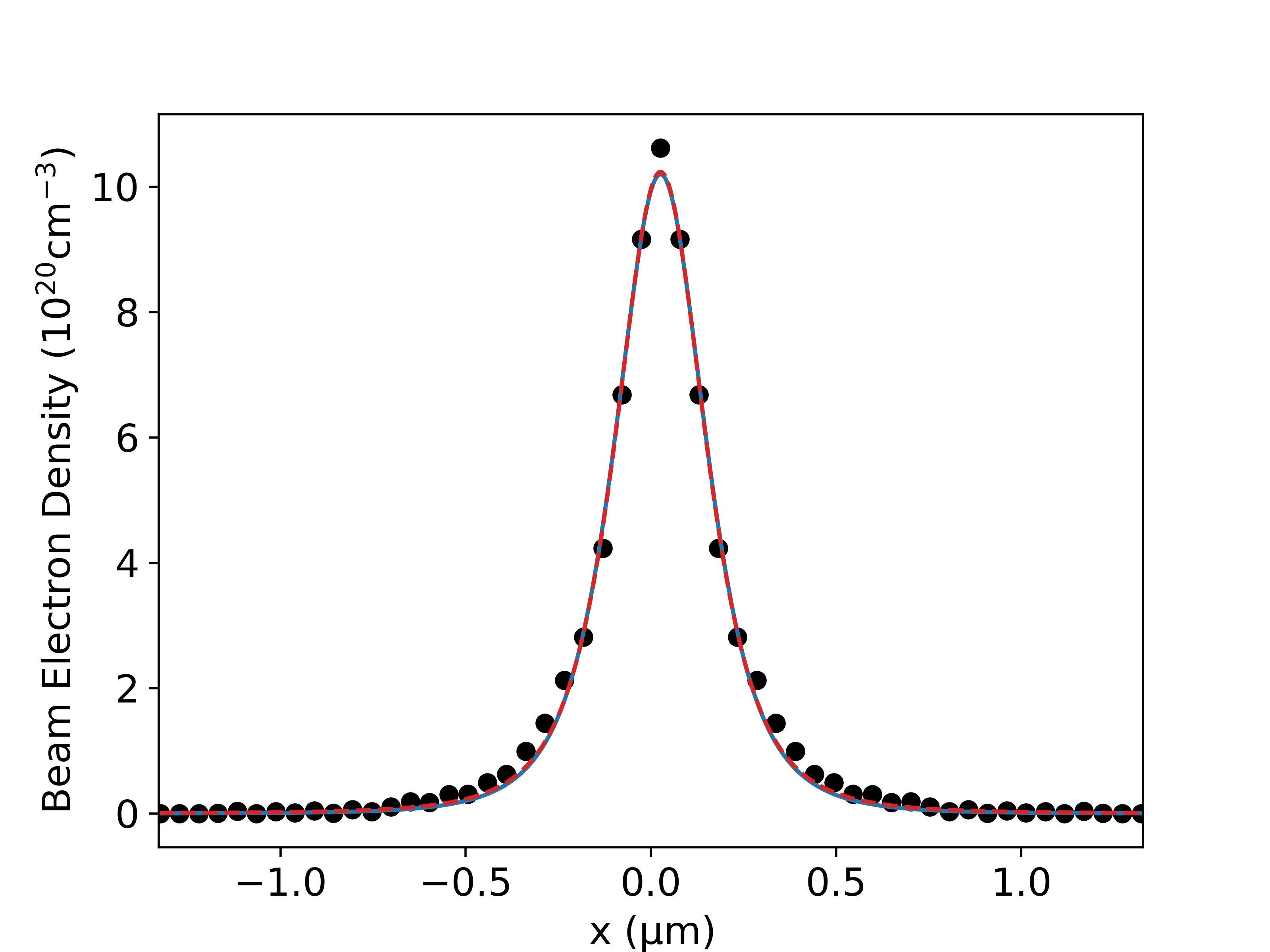}
\caption{}    
\end{subfigure}
\begin{subfigure}[h]{\columnwidth}
\centering
\includegraphics[width=\columnwidth]{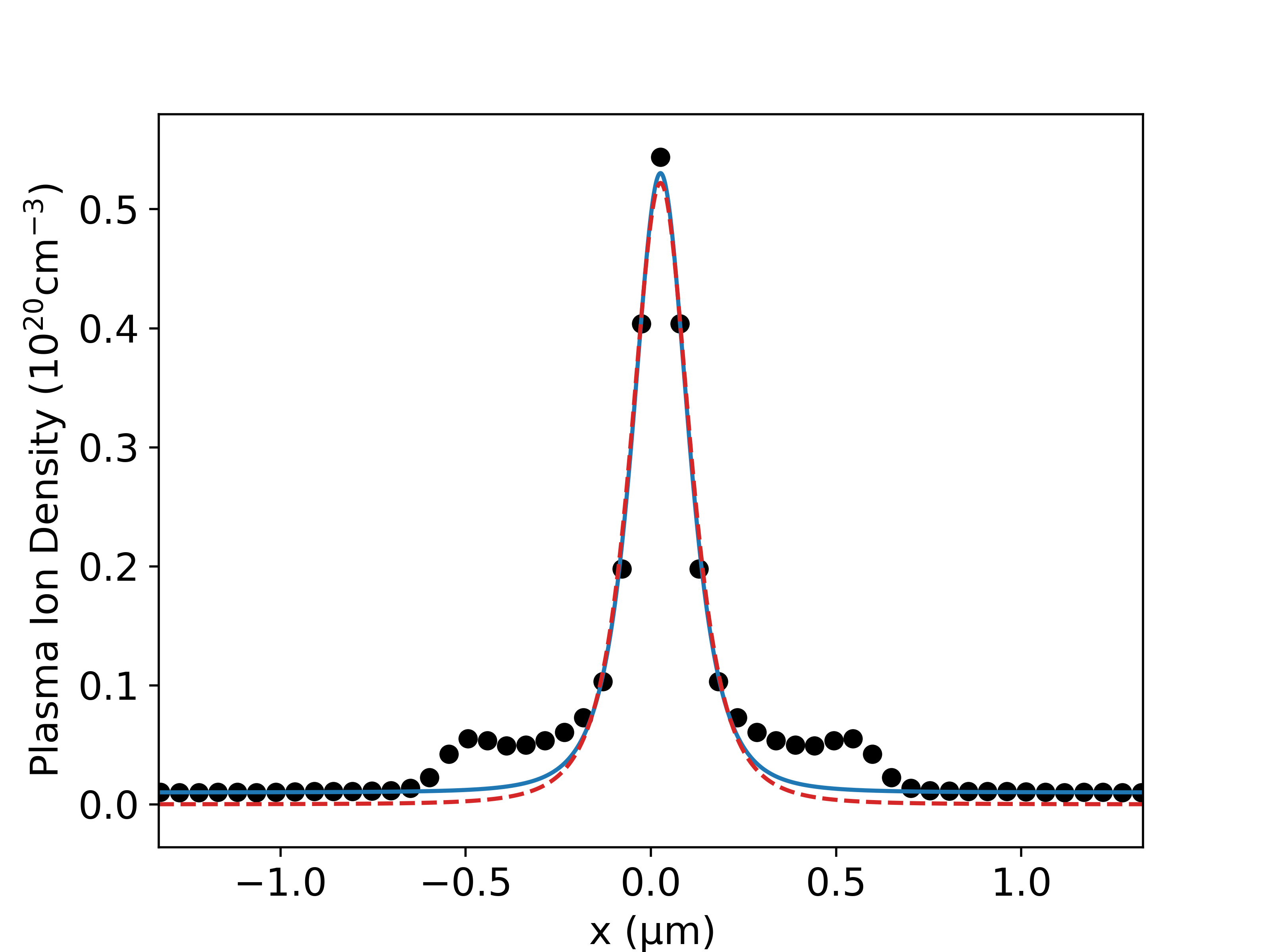}
\caption{}
\end{subfigure}
\caption{Density lineout plot of beam electrons (a) and plasma ions (b) for $y = 0$ and $\xi = ct - z = 150 \si{\mu m}$ at the end of the ion collapse simulation. Shown are the PIC simulation data (black dots), a fit to the unmodified Bennett profile Eqs.~(\ref{eq:ebennett}, \ref{eq:ibennett}) (dashed red line), and a fit to the modified Bennett profile Eqs.~(\ref{eq:modifiedibennett}, \ref{eq:modifiedebennett}) (solid blue line). Fits were performed using the nonlinear least-squares method to fit the modified and unmodified Bennett profiles to the PIC simulation data. Simulation performed using QuickPIC with the parameters shown in Table \ref{tab:colsimparams}.}
\label{fig:collapse}
\end{figure*}

\begin{figure*}
    \centering
    \begin{subfigure}[b]{0.475\textwidth}
        \centering
        \includegraphics[width=\textwidth]{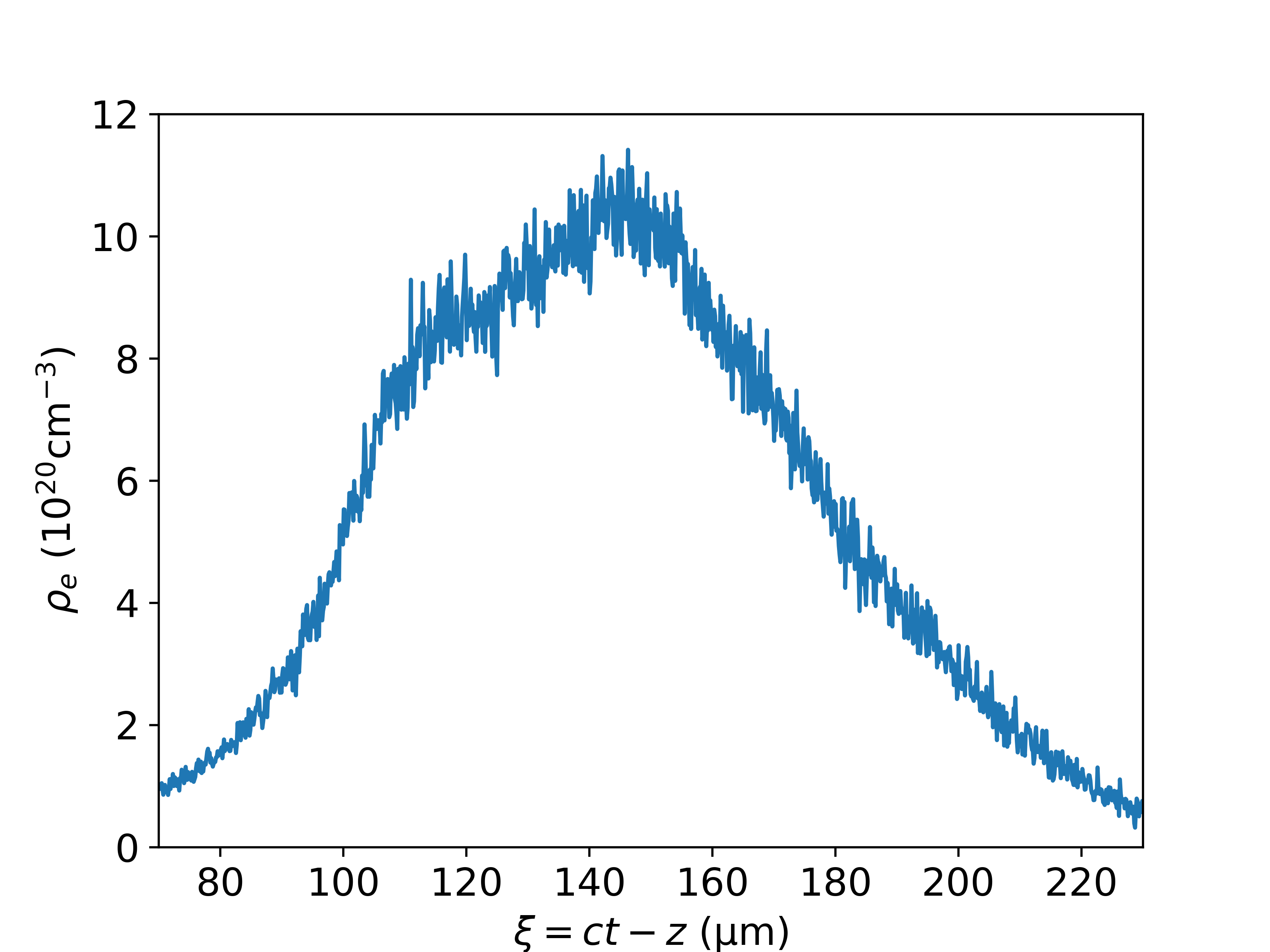}
        \caption{}  
    \end{subfigure}
    \hfill
    \begin{subfigure}[b]{0.475\textwidth}  
        \centering 
        \includegraphics[width=\textwidth]{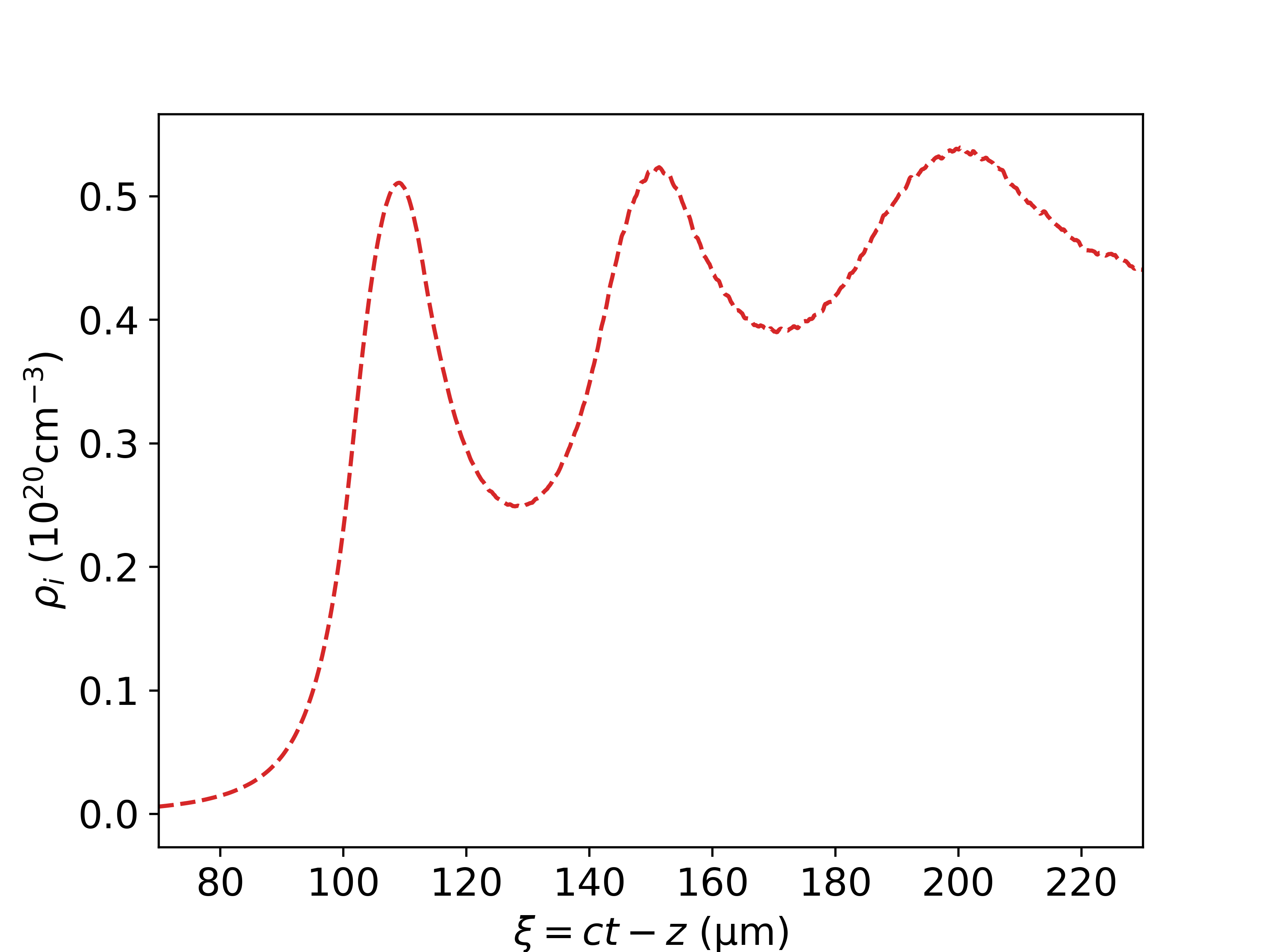}
        \caption{}
    \end{subfigure}
    \vskip\baselineskip
    \begin{subfigure}[b]{0.475\textwidth}   
        \centering 
        \includegraphics[width=\textwidth]{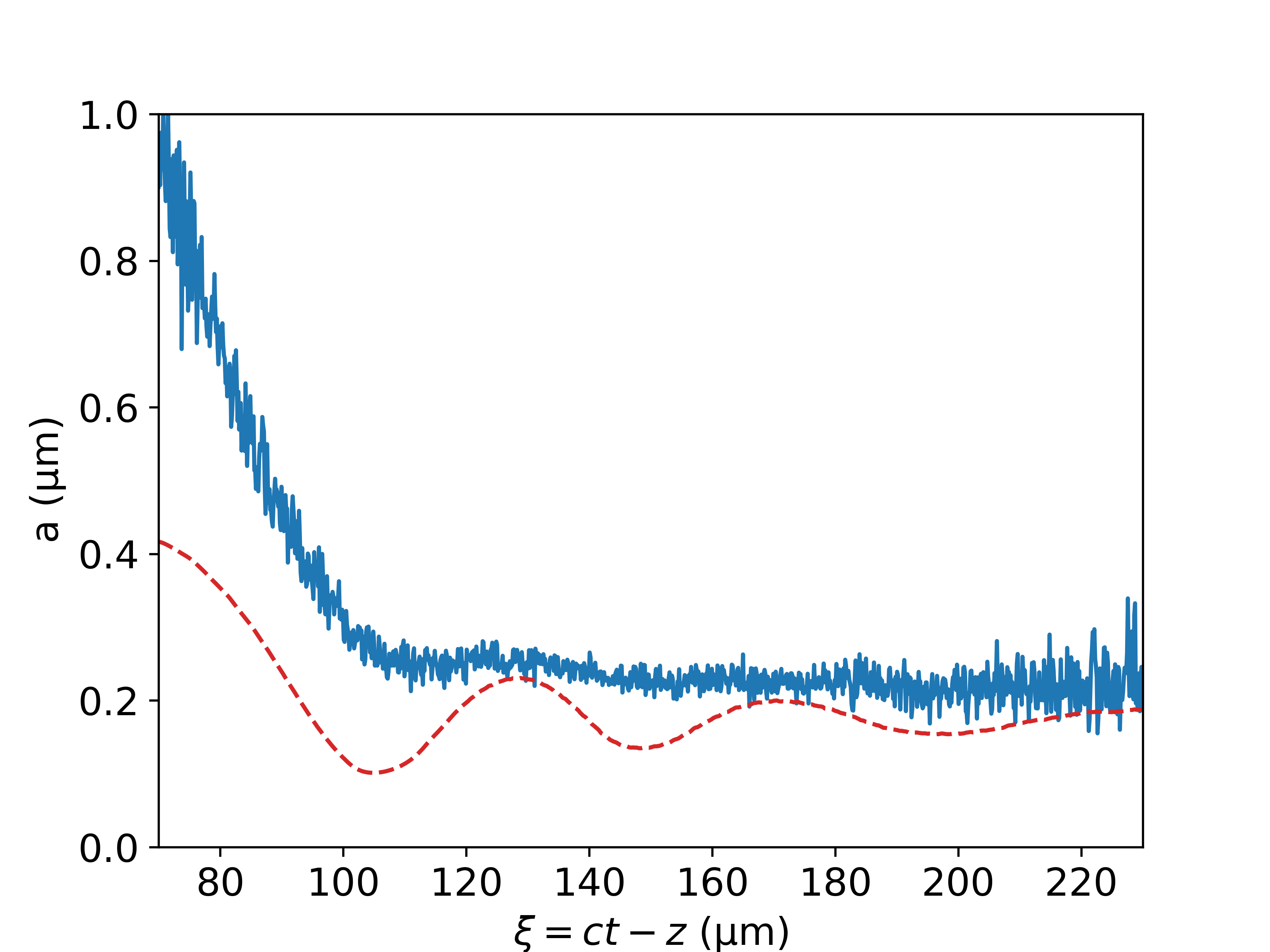}
        \caption{}
    \end{subfigure}
    \quad
    \caption{For each value of $\xi = ct - z$ of the beam electron and plasma ion densities at $y = 0$ at the end of the ion collapse simulation, a modified Bennett profile Eqs.~(\ref{eq:modifiedibennett}, \ref{eq:modifiedebennett}) was fit to the transverse density lineout using the nonlinear least-squares method. The three parameters $\rho_e$, $\rho_i$, and $a$ of these fits are plotted verses $\xi$ in (a), (b), and (c) respectively. Solid blue lines are fits of the beam electron density to the parameters in  Eq.~($\ref{eq:modifiedebennett}$), while dashed red lines are fits of the plasma ion density to the parameters in Eq.~($\ref{eq:modifiedibennett}$). Data from positions in $\xi$ are only shown from $70 \si{\mu m}$ to $230 \si{\mu m}$ as the beam density outside this range is negligible and the parameters of the fit to the beam density are not meaningful.}
    \label{fig:collapsefitparameters}
\end{figure*}

PIC simulations were performed in order to verify that an approximate equilibrium is reached in which the beam electron and plasma ion transverse densities are described by a Bennett-type profile after ion collapse. Simulations were performed using QuickPIC, a collisionless 3D parallel quasi-static particle-in-cell code \cite{qpic1, qpic2}. Because the collapse to near-equilibrium is very fast compared to the full time scale of the PWFA interaction, we use the PIC simulations only to examine this short (few 10's of femtosecond) transient period, leaving longer time-scale issues such as emittance growth due to scattering after near equilibrium has been reached for the tracking code studies described below. Similar to the approach of \cite{an}, the ion channel is created prior to the arrival of the beam involved in the collapse dynamics. 

\begin{table}[h]
    \centering
    \begin{tabular}{|c|c|c|}
        \hline
        Parameter & Value & Unit \\
        \hline
        $n_0$ & $10^{18}$ & $\si{cm^{-3}}$ \\
        $E$ & $10$ & $\si{GeV}$ \\
        $k_p^{-1}$ & $5.31$ & $\si{\mu m}$ \\
        $l_{\text{plasma}}$ & $10$ & $\si{cm}$ \\
        $\sigma_{x, \text{witness}}$ & $0.326$ & $\si{\mu m}$ \\
        $\sigma_{y, \text{witness}}$ & $0.326$ & $\si{\mu m}$ \\
        $\sigma_{p_x, \text{witness}}$ & $6.13$ & $m_e c$ \\
        $\sigma_{p_y, \text{witness}}$ & $6.13$ & $m_e c$ \\
        $\sigma_{z, \text{witness}}$ & $40$ & $\si{\mu m}$ \\
        $Q_{\text{witness}}$ & $2.89$ & $\si{nC}$ \\
        witness peak density & $280$ & $n_0$ \\
        \hline
    \end{tabular}
    \caption{The beam and plasma parameters of the ion collapse simulation.}
    \label{tab:colsimparams}
\end{table}The beam and plasma parameters of these ion collapse PIC simulation are given in Table \ref{tab:colsimparams} and are based on what is likely achievable in FACET-II experiments with beam optics solutions using permanent magnetic quadrupoles \cite{jlim}. The plasma density and bunch length are chosen to give ample interaction time for near-equilibrium to develop, with  an estimate of $k_i\sigma_z\simeq 4.2$. After the collapse, the value of $k_i\sigma_z$ is increased further, to above 25, as seen from the results indicated below. 

While we have developed a theoretical framework to understand the dual Bennett-type equilibrium in this system, the context of this present work is also experimentanl. In this regard, the final matching of the beam to the extremely strong ion focusing after collapse is experimentally very challenging in the FACET-II context due to final focus limitations. As such, in the simulations the  initial beam spot size was matched to the linear focusing scenario, a condition that is experimentally in reach.  This of course implies that there will be emittance growth present in this simulation which can be avoided with more careful matching. Because the beam is not \textit{nonlinearly} matched, the values of $\rho_e$, $\rho_i$, and $a$ in the simulation cannot be compared to the theoretical expressions in Eqs.~(\ref{eq:rhoeexp}), (\ref{eq:rhoiexp}), and (\ref{eq:bennettexp}) -- one may point to their consistency given the degree of observed emittance growth. Further,  matching the beam to mitigate emittance growth during ion collapse using the method developed in Ref.~\cite{an} requires a very time-intensive optimization, for the simulation. As such, the comparison to nonlinear matched simulations awaits a future investigation undertaken with enhanced computational resources. 

The nearly equilibrated beam electron and plasma ion densities inside the PWFA bubble at the end of the simulation relevant to FACET-II are shown in Fig.~\ref{fig:heatmap}. The beam and ion density line-out profiles at $\xi = 150 \si{\mu m}$, The ion density line-out clearly displays the oscillation period expected, near 60 $\mu$m. This figure also shows the fits of the radial distribution functions of the ions and electrons to both the unmodified Bennett profile discussed in Sec.~(\ref{subsec:bennett}) and the modified Bennett profile discussed in Sec.~(\ref{subsec:modifiedbennett}).  The fits were performed using nonlinear least squares fitting. The parameters of the fit to modified Bennett profiles as functions of $\xi$ are shown in Fig.~\ref{fig:collapsefitparameters}. 

It can be seen from these figures that the beam electron distribution resembles both the unmodified and modified Bennett distribution, which are nearly identical. The plasma ion distribution, unlike the beam electron distribution, displays some notable deviations from the (modified) Bennett equilibrium. This is because, even though we have not perfectly matched the beam to the final size, the beam electrons begin the interaction as a spatially gaussian, thermal distribution, which is quite similar to a Bennett distribution up to a relatively modest difference in spatial extent due to use of the linear matching conditions.  The initial state of the plasma ions is a cold, uniform distribution, however, which is very far from the near-equilibrium formed by fast collection (collapse) and nonlinear phase-mixing. It is observed in Fig.~\ref{fig:collapse} that the center of the ion distribution follows a modified Bennett profile fairly well while the residual effects of the non-thermal initial conditions are displayed as artifacts in the wings found at $r> a$. These artifacts can also be seen as a wake effect in Fig.~$\ref{fig:heatmap}$.

The maximum ion density after collapse in the simulation is found to be $\approx 50$ times the initial background ion density, a number consistent with theoretical predictions, which state that up a factor of 100 enhancement of density is expected in a nonlinear-matched scenario. The slightly smaller observed increase in density indicates that the final beam size is slightly increased due to phase space filamentation in the unmatched case that is to be experimentally encountered. Finally, we note that the large ion density present after collapse may give rise to significant emittance growth due to scattering, an issue we examine quantitatively below.

\section{Particle Tracking in Bennett Profile-Derived Fields} \label{sec:eqsim}

\subsection{Code Overview}

The modified Bennett equilibrium formed by the ions generates electrostatic fields with strongly nonlinear radial dependence. In addition, the ions are very dense, well in excess of atmospheric gas density. In order to establish the combined effects of these two phenomena encountered by beam electrons, simulations are utilized. The goal of these simulations is to quantify the emittance growth of the electron beam due to both multiple Coulomb scattering of beam electrons off of plasma ions and diffusion driven by plasma density fluctuations

Because emittance growth from these two sources take place on a timescale orders of magnitude longer than the ion collapse, a process already computationally intensive to simulate with PIC codes, it is not feasible to use a PIC approach to simulate these effects. Instead we have developed a custom 2D tracking code. When a simulation is run with this code, the initial positions and momenta of the beam electrons are randomly sampled from the modified Bennett distribution function. These electrons are then symplectically tracked through the analytically known electromagnetic fields which are derived assuming the system is in equilibrium. At the end of each step, for each particle, an angle is sampled from the scattering angle distribution predicted by Moli\`ere's theory, and the particle's momentum is deflected by this angle. Statistics describing the electron distribution are computed and saved at multiple $z$ positions for later analysis. Acceleration is accounted for  by adiabatically adjusting the beam energy and parameters which scale with the energy at each step. A sinusoidal plasma density modulation or gaussian noise in the ion density can be included to drive parametric resonances. The tracking simulation code is written in C++ and is parallelized using the Boost MPI library. Analysis and plotting is performed with Python 3. Simulations are run on UCLA's Hoffman2 computing cluster. The code is available on Github \footnote{Equilibrium Plasma Accelerator Multiple Scattering Simulator (EPAMSS): \url{https://github.com/clairehansel3/EPAMSS}.}.

\subsection{Tracking: Macroscopic Forces} \label{sec:trackingeqs}

As noted above, the PIC approach to understanding the behavior of the beam-plasma interaction is very useful for shorter time scales, but to study the longer term evolution of the beam, another approach based on tracking in using model Hamiltonian may be employed. This is analysis is intended to reveal slower, yet critically important phenomena such as phase space diffusion. We now derive the equations of motion used in the tracking code. To obtain the force on a beam electron, we begin with the modified plasma ion Bennett profile Eq.~(\ref{eq:modifiedibennett}). In order to explore parametric resonances we wish to vary $n_i$ as a function of $z$ in our dynamics model. This is done by multiplying $n_i$ by a modulation factor $m(z)$. The equations of motion are then easily obtained from Eq.~(\ref{eq:eforceapprox}). Introducing normalized variables: $\tilde{t} \equiv \omega_p t = \tilde{z} \equiv k_p z$, $\tilde{\vec{r}} \equiv k_p \vec{r}$, $\tilde{r} \equiv k_p r$, $\tilde{\rho_i} \equiv k_p^{-3} \rho_i$, and $\tilde{n_0} \equiv k_p^{-3} n_0$, and assuming as before ultra-relativistic ($\beta \simeq 1$) and paraxial ($p_{\bot} \ll p_z$) electron motion, the equations of motion are

\begin{equation} \label{eq:normeom}
\frac{d^2 \tilde{\vec{r}}}{d \tilde{z}^2} + \frac{1}{\gamma} \frac{d \gamma}{d \tilde{z}} \frac{d\tilde{\vec{r}}}{d \tilde{z}} +\frac{Z m(\tilde{z})}{2 \gamma}\left(1 + \frac{\frac{\rho_i}{n_0}}{1+ \left(\frac{\tilde{r}}{\tilde{a}}\right)^2} \right) \tilde{\vec{r}} = 0.
\end{equation}

\noindent Our intention is to use the symplectic velocity-Verlet method to solve the equations of motion numerically. However, due to the damping term in the above expression, the velocity-Verlet method cannot be employed. To remedy this we perform the coordinate transformation $\tilde{\vec{r}} = \gamma^{-\frac{1}{2}} \tilde{\vec{u}}$ which gives

\begin{equation}
\begin{split}
\frac{d^2 \tilde{\vec{u}}}{d \tilde{z}^2} &+ \left[ \frac{1}{4 \gamma^2} \left ( \frac{d \gamma}{d \tilde{z}} \right )^2 - \frac{1}{2 \gamma} \frac{d^2 \gamma}{d \tilde{z}^2} + \right . \\
&+ \left . \frac{Z m(\tilde{z})}{2 \gamma}  \left(1 + \frac{\frac{\tilde{\rho_i}}{\tilde{n_0}}}{1+ \frac{1}{\gamma} \left(\frac{\tilde{u}}{\tilde{a}}\right)^2} \right) \right] \tilde{\vec{u}} = 0
\end{split}
\end{equation}

which is now in the correct form to be solved numerically using the velocity-Verlet method. Constant acceleration is accounted for by adiabadically updating $\gamma$ and $a$ with $z$. Since $a$ scales as $\gamma^{-\frac{1}{4}}$, for constant acceleration we can write

\begin{equation}
\gamma(\tilde{z}) = \gamma_0 + \frac{d\gamma}{d\tilde{z}} \tilde{z}
\end{equation}

\noindent and 

\begin{equation}
\tilde{a}(\tilde{z}) = \tilde{a_0} \left(1 + \frac{1}{\gamma_0} \frac{d\gamma}{d \tilde{z}} \tilde{z} \right) ^ {-\frac{1}{4}}.
\end{equation}

\noindent where $\frac{d\gamma}{d\tilde{z}}$ is constant. Validation of the tracking algorithm is discussed in Appendix \ref{sec:trackingvalidation}. 

\subsection{Tracking: Coulomb Scattering} \label{sec:trackingcoulombscattering}

As the subject of relativistic particle Coulomb scattering in an ion environment is not commonly discussed, we discuss here the details of the computational approach to modeling the relevant scattering dynamics. The Coulomb scattering algorithm used in the tracking simulation is based on Moli\`ere's theory of small angle multiple scattering \cite{bethe, scott}. At the end of each computational step after a particle's phase space state has been tracked, a scattering angle is sampled from the scattering angle distribution which is given by Moli\`ere's theory. The velocity of the particle is then deflected by this angle. The step size and scattering angle are assumed to be small enough that the contribution of the scatter to the particle's position can be ignored. Because Moli\`ere's theory requires constant transverse density, it is assumed that the step size $\Delta z$ is small enough so that the plasma ion density along the particle's trajectory is constant. Note that Moli\`ere's theory assumes no ion recoil takes place during scattering.

Moli\`ere's original theory is based off a differential cross section which includes atomic screening, and so it must be modified to account for scattering off ions, which lack atomic screening. For this purpose, we use the Mott-Born differential cross section \cite{crosssections}.  Because the total cross section for the Mott-Born differential cross section diverges, a minimum angle $\theta_{\text{min}}$ is introduced and the cross section is assumed to vanish for $\theta < \theta_{\text{min}}$:

\begin{equation} \label{eq:diffcrosssec}
\frac{d\sigma}{d\Omega}(\theta) = 
\begin{cases}
    \frac{Z^2 r_e^2 (1 - \beta_0^2\sin^2(\frac{\theta}{2}))}{4 \gamma_0^2 \beta_0^4 \sin^4(\frac{\theta}{2})} & \theta \geq \theta_{\text{min}},\\
    0 & \theta < \theta_{\text{min}}.
\end{cases}
\end{equation}

The code computes $\theta_{\text{min}}$ from the total cross section $\sigma$ which is defined as $\sigma \equiv \pi r_\sigma^2$ where the cross sectional radius $r_\sigma$ is an adjustable parameter of the simulation. Modifying the Moli\`ere theory by using the Mott-Born differential cross section (Eq.~\ref{eq:diffcrosssec}) gives the scattering angle distribution, with the following relations applied:

\begin{equation} \label{eq:distsum}
f(\vartheta) = 2\vartheta e^{-\vartheta^2} + \sum_{n=1}^{\infty}\frac{f^{(n)}(\vartheta)}{B^n};
\end{equation}

\begin{equation} \label{eq:scaryintegral}
f^{(n)}(\vartheta) = \frac{\vartheta }{n!}\int_0^{\infty} e^{\frac{-\eta^2}{4}}\left( \frac{\eta^2}{4} \ln \left( \frac{\eta^2}{4} \right) \right)^n J_0(\eta \vartheta) \eta d\eta;
\end{equation}

\begin{equation}
B = -W_{-1}\left( \frac{-e^{-2(1 - \gamma_E)}}{\Omega_0} \right);
\end{equation}

\begin{equation} \label{eq:thfromvarth}
\theta = \left(\theta_{\text{min}} \sqrt{\Omega_0 B} \right) \vartheta;
\end{equation}

\begin{equation}
\theta_{\text{min}} = \frac{2 Z r_e}{\gamma_0 \beta_0^2 r_\sigma};
\end{equation}

\begin{equation}
\Omega_0 = n_i \sigma \Delta z.
\end{equation}

Here $W_{-1}(x)$ is the lower branch of the Lambert W-function \cite{lambertw}, $\gamma_E$ is the Euler-Mascheroni constant, and $n_i$ is the density of plasma ions at the position of the particle, $\Delta z$ is the step size of the simulation, and $Z$ is the ion atomic number. The function $f(\vartheta)$ is normalized such that $\int_0^{\infty} f(\vartheta) d\vartheta = 1$. The results stated above are derived in Appendix \ref{sec:multiplescattering}. Before a simulation is initiated, a table of values for $f^{(n)}(\vartheta)$ is created by evaluating this function at evenly spaced points between zero and a cutoff value $\vartheta_{max}$. The evaluation is performed by using an adaptive Gauss-Kronrod quadrature to numerically integrate (\ref{eq:scaryintegral}). This permits  $f^{(n)}(\vartheta)$ to be evaluated during the simulation by using cubic B-spline interpolation on the stored tables. The simulation truncates the sum in Eq.~\ref{eq:distsum} at a maximum order which is specified as a simulation parameter. In order to evaluate scattering at an angle $\theta$, the code samples the scaled angle $\vartheta$ from Eq.~\ref{eq:distsum} using rejection sampling on an initial uniform distribution from $0$ to $\vartheta_{max}$, and then the angle $\theta$ is computed using Eq.~(\ref{eq:thfromvarth}). If the particle is far enough off axis, the number of scatters is small enough that Moli\`ere's theory does not apply. As discussed in Appendix \ref{sec:multiplescattering}, Moli\`ere's theory requires $\Omega_0 \gtrsim 25$. If a particle is far enough off-axis such that this condition is not met, the effects of scattering for that particle are assumed to be negligible and ignored. To determine the effects of scattering, the simulation is run twice with the same starting particles, once with scattering turned on and once with scattering off. The validation of the scattering algorithm is discussed in Appendix \ref{sec:scatteringvalidation}.

\subsection{Tracking: Results} \label{sec:tcresults}

\begin{table}[]
    \centering
    \begin{tabular}{|c|c|c|}
        \hline
        Parameter & Value & Unit \\
        \hline
        Species & H & - \\
        $E$ & $10$ & $\si{GeV}$ \\
        Acceleration gradient & $10$ & $\si{GeV/m}$ \\
        $l_{\text{plasma}}$ & $100$ & $\si{m}$ \\
        $n_0$ & $10^{18}$ & $\si{cm^{-3}}$ \\
        $\rho_i$ & $10^{20}$ & $\si{cm^{-3}}$ \\
        $a$ & $170$ & $\si{nm}$ \\
        $r_\sigma$ & $10$ & $\si{nm}$ \\
        $\vartheta_{\text{max}}$ & 10 & - \\
        Simulation particles & $400000$ & - \\
        Maximum $f(\vartheta)$ order & 3 & - \\
        Minimum steps per betatron period & 200 & - \\
        Numerical integration tolerance & $10^{-10}$ & - \\
        Maximum integration recursions & 15 & - \\
        Spline points & 1000 & - \\
        \hline
    \end{tabular}
    \caption{Parameters of the FACET-II simulation.}
    \label{tab:eqsimfacetparams}
\end{table}

\begin{table}[]
    \centering
    \begin{tabular}{|c|c|c|}
        \hline
        Parameter & Value & Unit \\
        \hline
        Species & H & - \\
        $E$ & $10$ & $\si{GeV}$ \\
        Acceleration gradient & $10$ & $\si{GeV/m}$ \\
        $l_{\text{plasma}}$ & $100$ & $\si{m}$ \\
        $n_0$ & $10^{18}$ & $\si{cm^{-3}}$ \\
        $\rho_i$ & $10^{20}$ & $\si{cm^{-3}}$ \\
        $a$ & $25$ & $\si{nm}$ \\
        $r_\sigma$ & $10$ & $\si{nm}$ \\
        $\vartheta_{\text{max}}$ & 10 & - \\
        Simulation particles & $400000$ & - \\
        Maximum $f(\vartheta)$ order & 3 & - \\
        Minimum steps per betatron period & 200 & - \\
        Numerical integration tolerance & $10^{-10}$ & - \\
        Maximum integration recursions & 15 & - \\
        Spline points & 1000 & - \\
        \hline
    \end{tabular}
    \caption{Parameters of the LC simulation.}
    \label{tab:eqsimlcparams}
\end{table}

\begin{figure}[htb]
\centering
\includegraphics[width=\columnwidth]{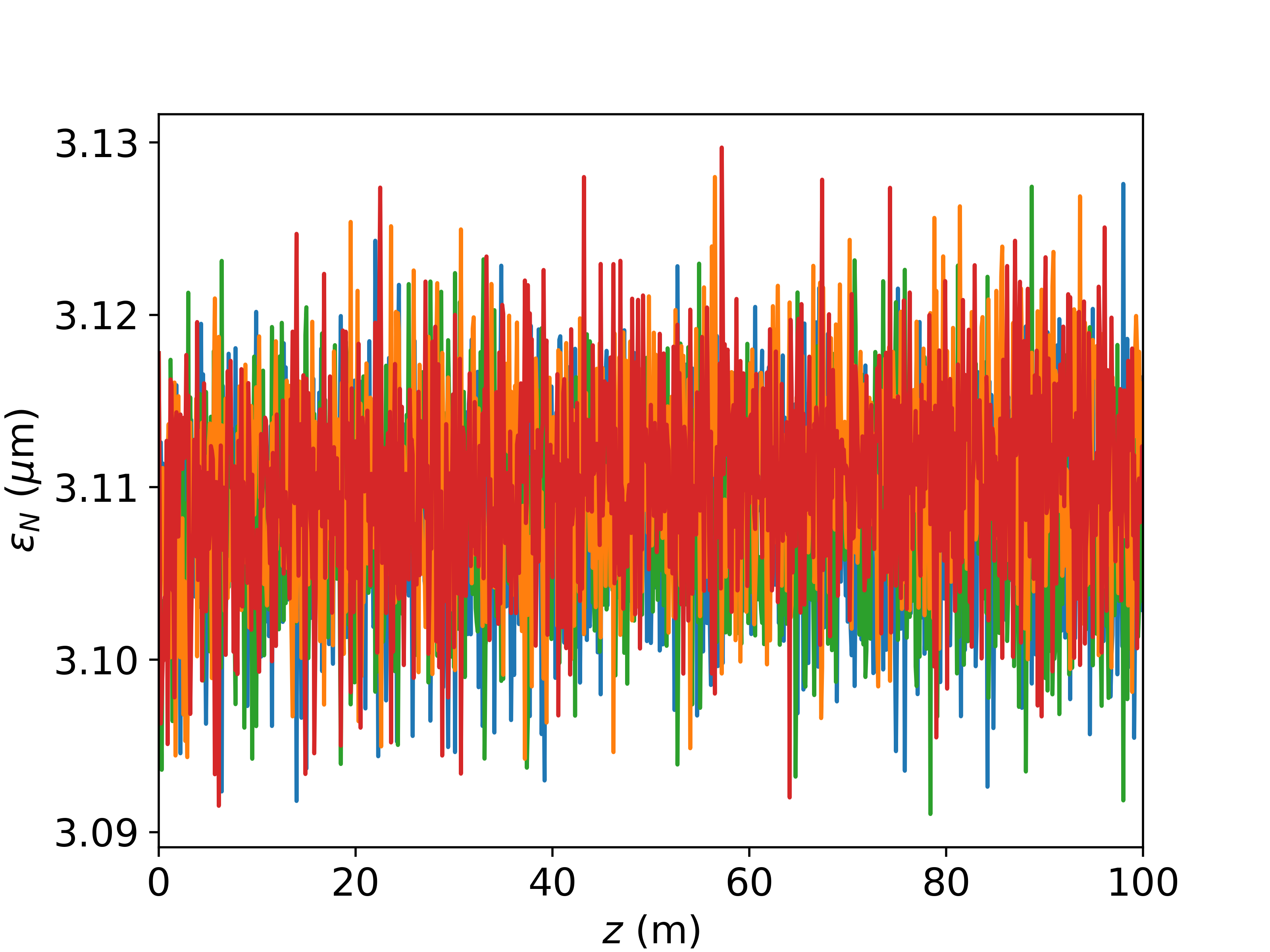}
\caption{Simulated growth of normalized emittance due to scattering for the FACET-II case. The orange and red lines show the $x$ and $y$ emittances, respectively, when scattering is included while the blue and green lines show the $x$ and $y$ emittances respectively when scattering is not included. The simulation was run with the parameters shown in Table \ref{tab:eqsimfacetparams}.}
\label{fig:facetemit}
\end{figure}

\begin{figure}[htb]
\centering
\includegraphics[width=\columnwidth]{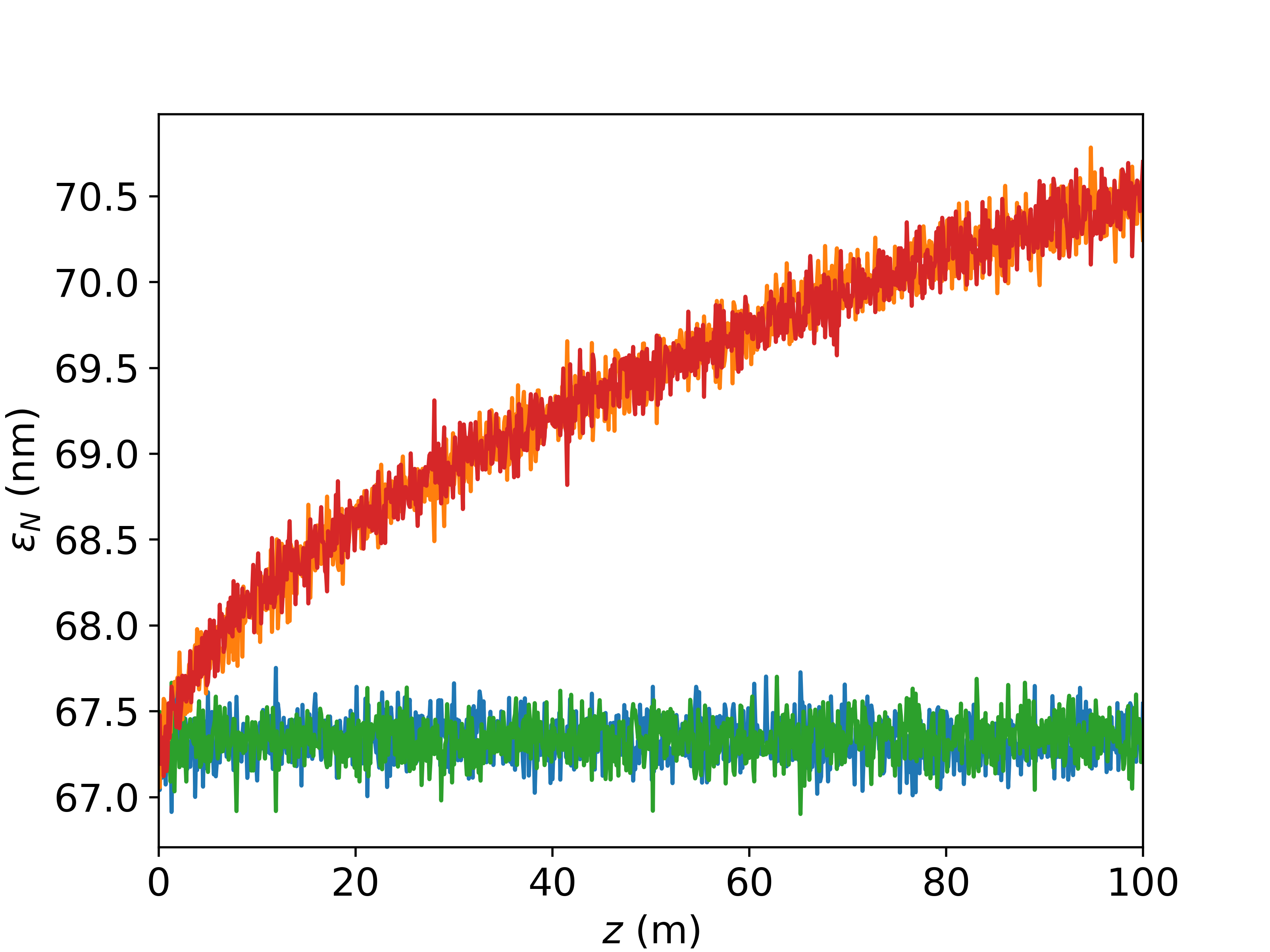}
\caption{Simulated growth of normalized $x$ emittance due to scattering for a hypothetical linear collider. The orange and red lines show the $x$ and $y$ emittances, respectively, when scattering is included, while the blue and green lines show the $x$ and $y$ emittances, respectively, when scattering is not included. The simulation was run with the Parameters shown in Table \ref{tab:eqsimlcparams}.}
\label{fig:lcemit}
\end{figure}

The tracking code was first run with parameters based on what is achievable at FACET-II, although the plasma was simulated to be $100 \si{m}$ to have more distance to permit observation of possible emittance growth. The tracking run parameters are shown in Table \ref{tab:eqsimfacetparams}. The observed emittance growth is plotted in Fig.~\ref{fig:facetemit}. From this plot it is clear that the emittance growth due to scattering in this particular case is negligible.

The tracking code was then run with parameters associated with a hypothetical LC. These parameters are shown in Table \ref{tab:eqsimlcparams}. The emittance growth is plotted in Fig.~\ref{fig:lcemit}. From this plot we can see that the emittance growth in this case is small but non-negligible.

\section{Chaotic Behavior and Parametric Resonances}
\label{sec:stability}

While scattering is of primary importance in giving phase space diffusion and emittance growth, the nonlinearity of the restoring force can give rise to other relevant processes. In particular, one may excite parametric resonances due to periodic changes in the focusing ion distribution population from plasma density fluctuations or breathing of the bulk beam-ion distribution. In this regard, we note that there is a range of effective locally resonant frequencies due to the non-uniformity of the ion distribution, ranging from $\sqrt{2 \pi r_e Z n_0 / \gamma}$ at large amplitude, to a near-axis frequency of $\sqrt{2 \pi r_e Z (\rho_i + n_0) / \gamma}$. In order to evaluate possible phase space diluting effects, we have again employed the tracking code.

\begin{figure*}
    \centering
    \begin{subfigure}[b]{0.475\textwidth}
        \centering
        \includegraphics[width=\textwidth]{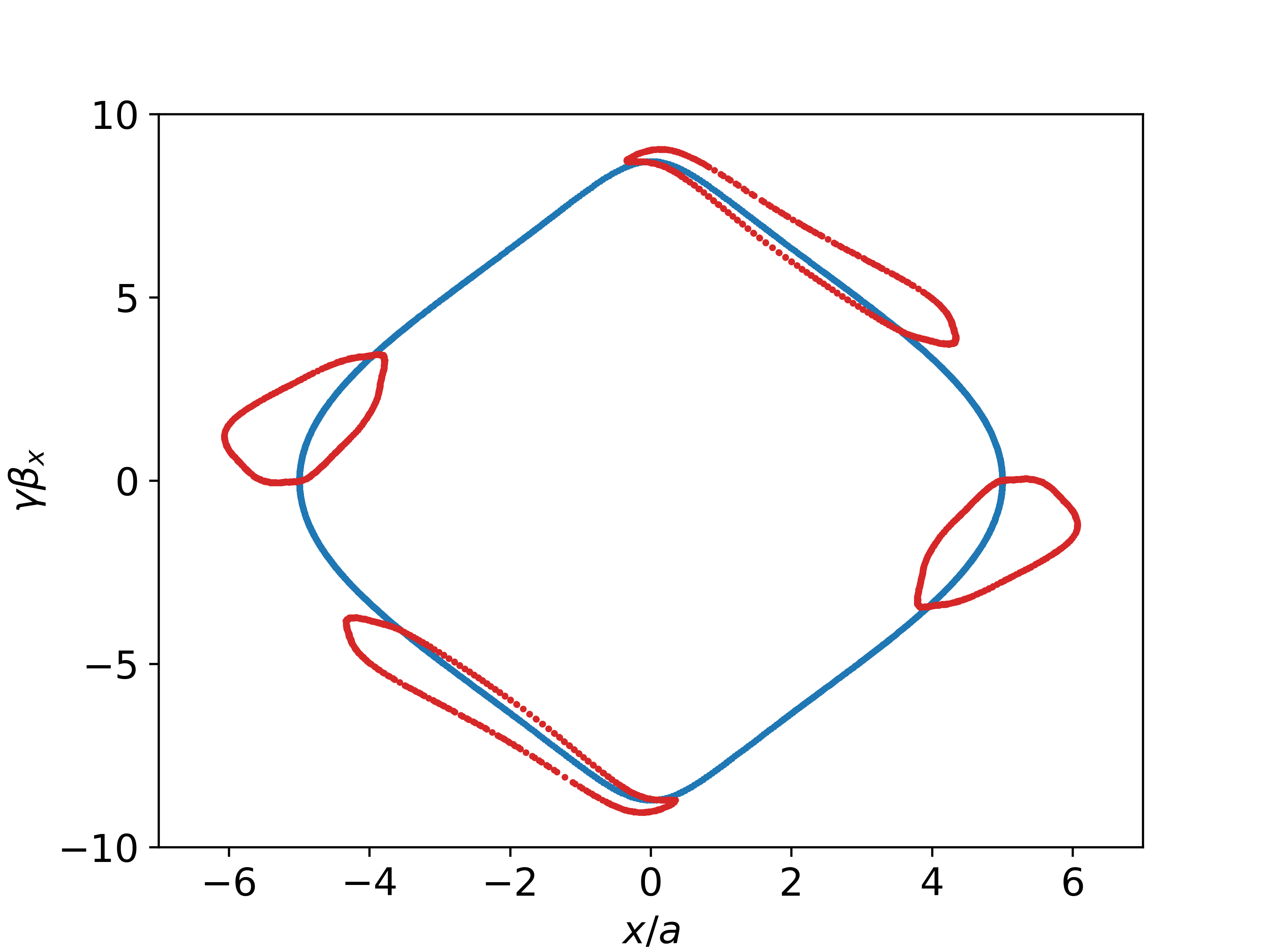}
        \caption{}  
    \end{subfigure}
    \hfill
    \begin{subfigure}[b]{0.475\textwidth}  
        \centering 
        \includegraphics[width=\textwidth]{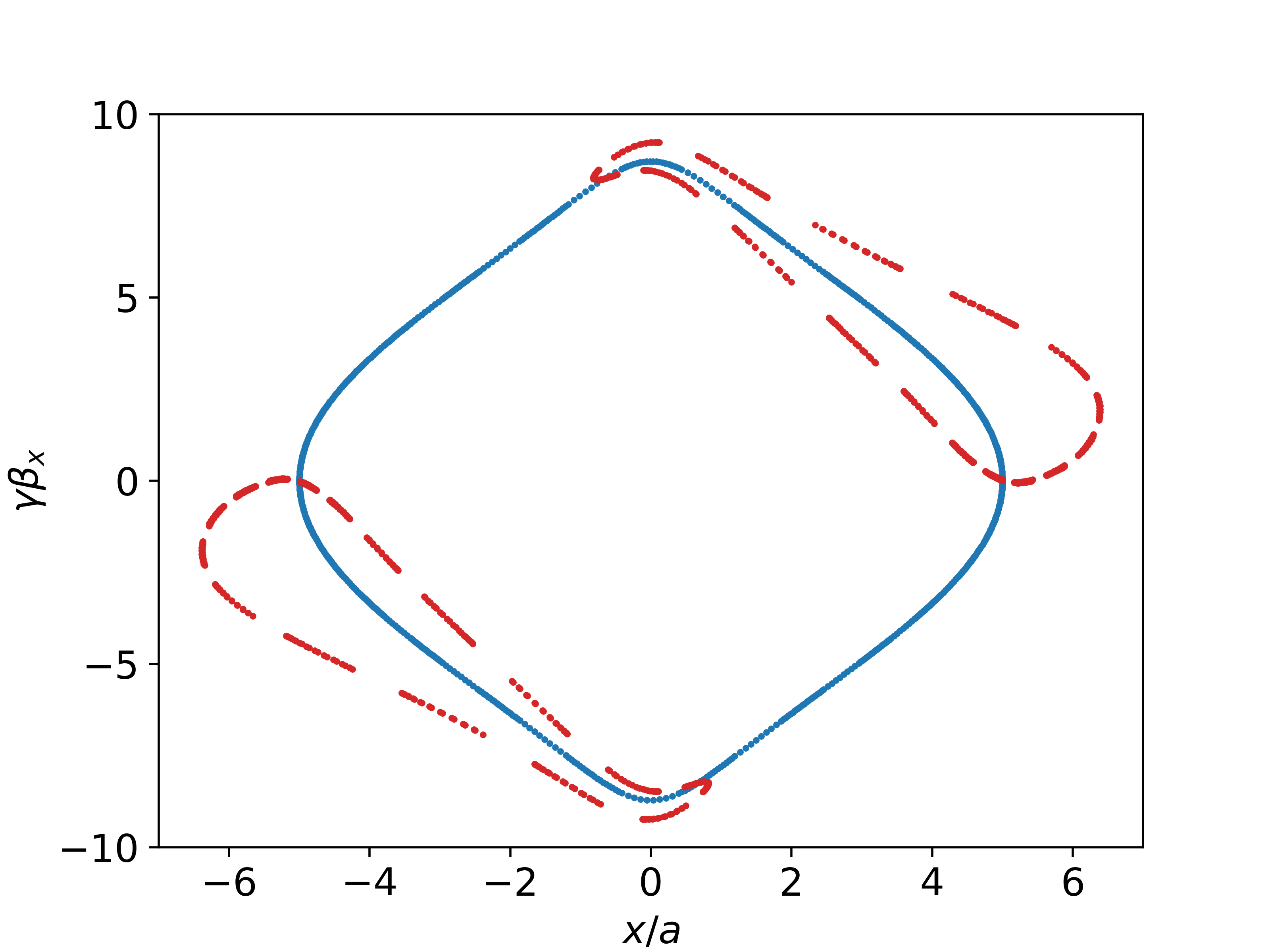}
        \caption{}
    \end{subfigure}
    \vskip\baselineskip
    \begin{subfigure}[b]{0.475\textwidth}   
        \centering 
        \includegraphics[width=\textwidth]{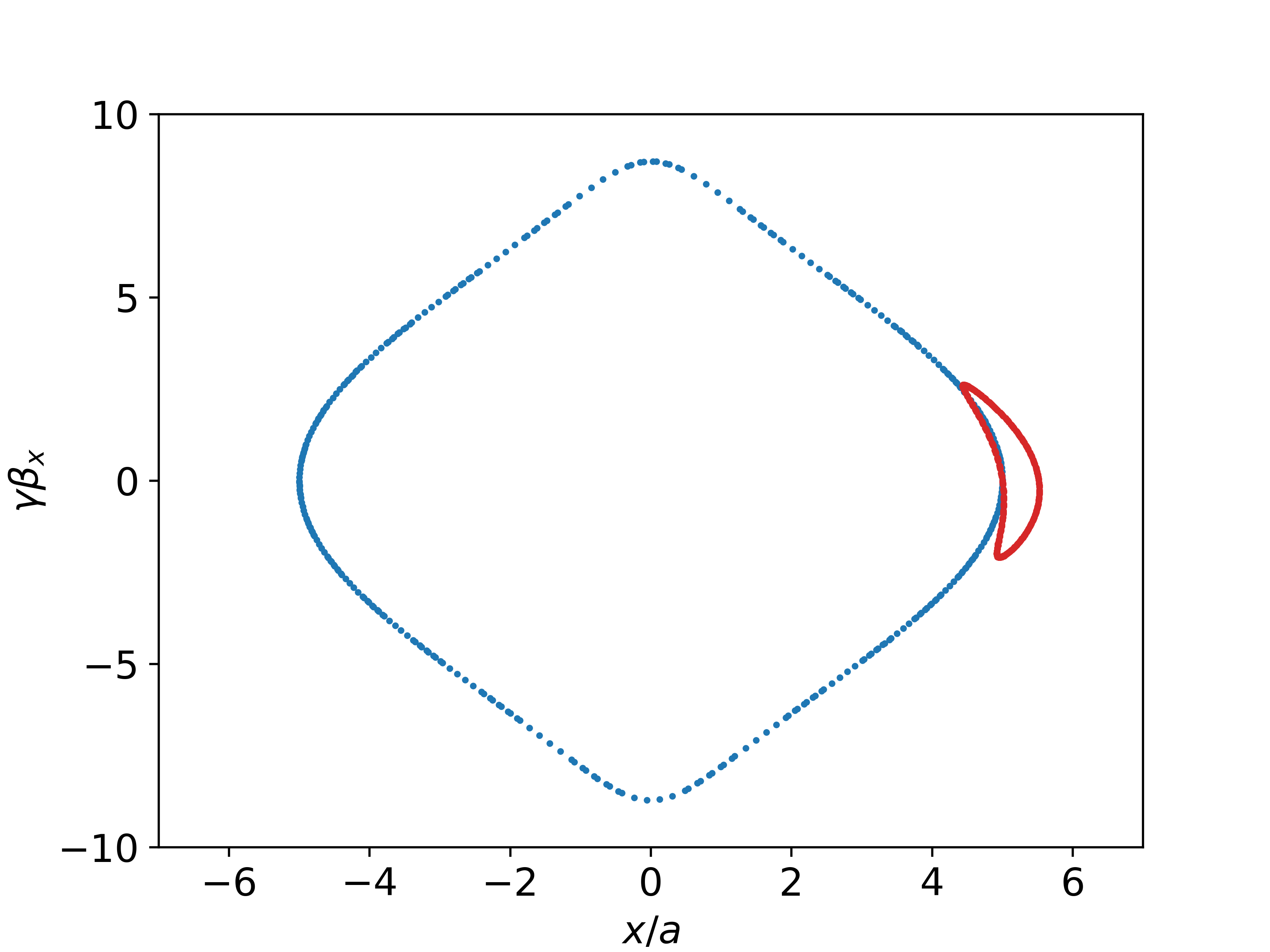}
        \caption{}
    \end{subfigure}
    \quad
    \caption{
    Poincar\'e plots showing parametric resonances excited by multiplying the plasma ion density by a longitudinal modulation factor $m(z) = 1 + A_{\textup{modulation}} \sin(k_{\textup{modulation}} z)$. Poincar\'e plot data are sampled at a frequency equal to the modulation frequency. Three plots are shown: $k_{\textup{modulation}} = k_{\beta, r \ll a}$ (a), $k_{\textup{modulation}} = k_{\beta, r \ll a} / 2$ (b), and $k_{\textup{modulation}} = k_{\beta, r \ll a} / 4$ (c), where $k_{\beta, r \ll a}$ is the on axis betatron angular wavenumber. Blue dots indicate the modulation is turned off ($A_{\textup{modulation}} = 0$) while red dots indicate the modulation is turned on ($A_{\textup{modulation}} = 0.1$). This simulation was run with the LC parameters shown in Table \ref{tab:eqsimlcparams} except that a plasma length of $1 \si{m}$ was used and only a single particle was tracked.}
    \label{fig:poincare}
\end{figure*}

To demonstrate resonances and island formation, simulations were performed in which a sinusoidal modulation was added to the ion density. This was done by employing the modulation function $m(z)$ as defined in Sec.~(\ref{sec:trackingeqs}) to

\begin{equation}
m(z) = 1 + A_{\textup{modulation}} \sin(k_{\textup{modulation}} z).
\end{equation}

For these simulations, a single particle was tracked through phase space. This particle was initialized at $x = 5a$, $y =0$ and $p_x = p_y = 0$ at the beginning of the plasma. Simulations were performed with the scattering process turned off and with the LC parameters shown in Table \ref{tab:eqsimlcparams} except $l_{\textup{plasma}} = 1 \si{m}$ was used. The relative amplitude of the modulation $A_{\textup{modulation}}$ was chosen to be $0.1$. Poincare plots were created by sampling the phase space coordinates at values of $z$ which are multiples of the near axis betatron wavelength. Poincar\'e plots obtained from these tracking simulations show island formation in cases where the modulation wave-numbers $k_{\textup{modulation}}$ were equal to or near $1$, $1/2$, or $1/4$ times the near axis betatron wavenumber. The Poincar\'e plots for these three cases are shown in Fig.~\ref{fig:poincare}.

\begin{figure}[htb]
\centering
\includegraphics[width=\columnwidth]{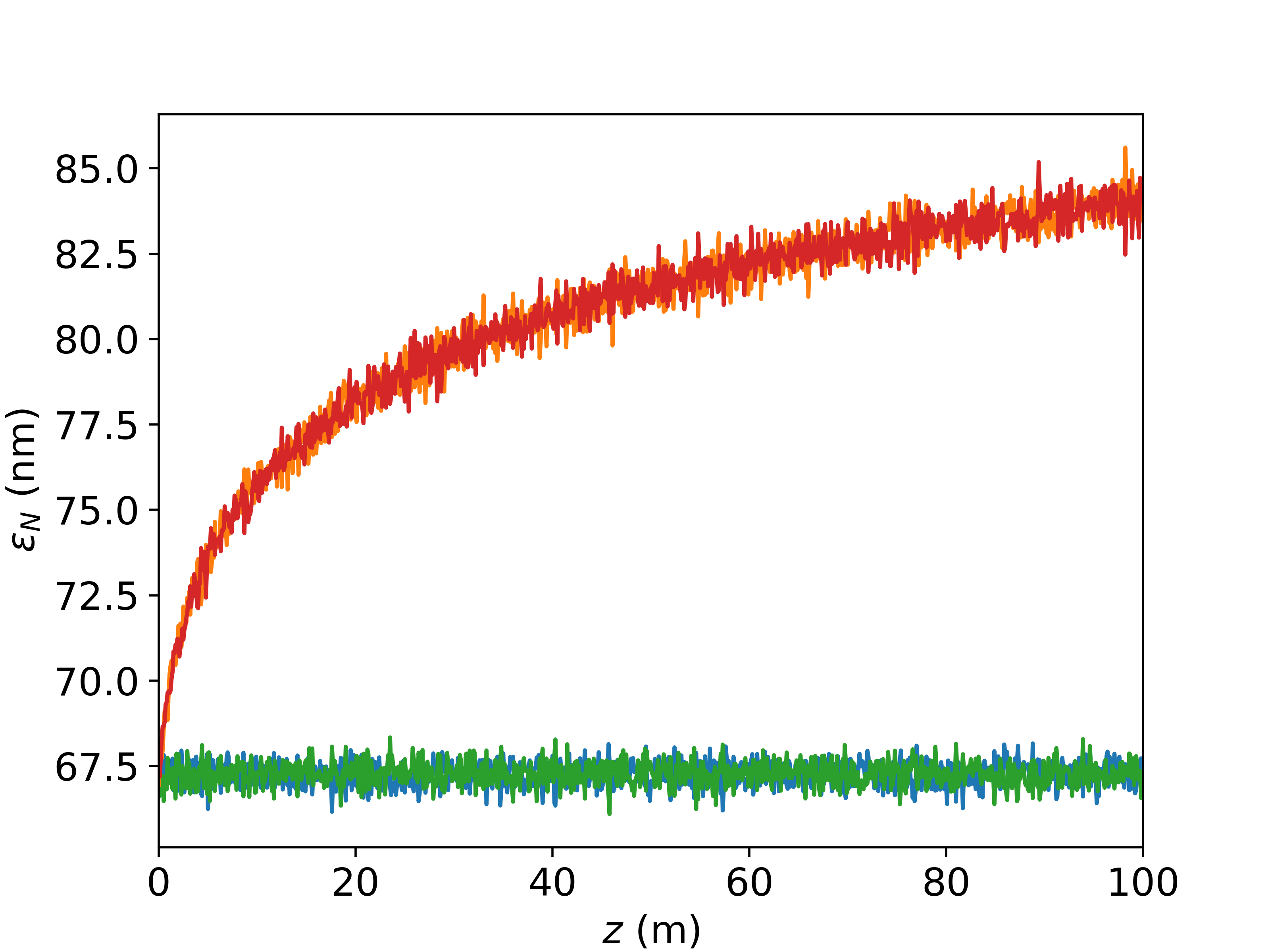}
\caption{Simulated growth of normalized $x$ emittance due to ion density white noise. The orange and red lines show the $x$ and $y$ emittances, respectively, when a 1\% gaussian white noise is added to the ion density, while the blue and green lines show the $x$ and $y$ emittances, respectively, when no noise was added. Effects due to scattering or sinusoidal density modulation were not included in this simulation. This simulation was run with the LC parameters shown in Table \ref{tab:eqsimlcparams}, except only 50,000 particles were tracked.}
\label{fig:gaussemit}
\end{figure}

In order to provide a physical seed for observing diffusive emittance growth due to excitation of parametric resonances, white noise was added to the ion density. This was done by at each step $z_i$ sampling from a gaussian with $\sigma = 0.01$. This sampled value, added to unity, was used as the value of $m(z_i)$. The LC simulation from Sec.~(\ref{sec:tcresults}) was repeated without scattering or sinusoidal modulation. The emittance growth is this scenario is shown in Fig.~\ref{fig:gaussemit}; it is significant but not unmanageably large. It is of similar importance as the scattering-induced diffusive emittance growth. In the future, a Fokker-Planck approach might be used to better understand diffusion.

\section{Conclusions and Outlook} \label{sec:conclusion}

In this paper we added to the analytical understanding of nonlinear equilibria in PWFAs with ion motion. We utilized and extended previous research, using both theoretical and computational approaches, to confirm that these equilibria are described by Bennett-type profiles, and given an estimate of the associated dimensions such as the Bennett radius $a$. We have, through tracking simulations, quantified the expected emittance growth due to scattering in a hypothetical PWFA based linear collider that includes strong ion motion and found it to be acceptably small. Finally we discussed diffusion due to parametric resonances in the nonlinear focusing fields of the Bennett-type profile, with similar conclusions reached considering possible problems encountered in LC scenarios.  These results thus significantly extend previous results obtained from  PIC simulations that demonstrated emittance mitigation for PWFAs with ion motion \cite{an}. The results of the current work give further reason to be optimistic in this regard.

The issue of ion motion has, due to the demanding conditions needed to explore the relevant physics, to date not been addressed experimentally. Our results give context which identifies the important physics to be investigated in firt experiments. Indeed, we have discussed above a scenario that is to be explored at FACET-II by the E-314 experimental collaboration. In this experiment, by utilizing cutting edge methods in photoinjector electron sources and beam preparation, in tandem with very high gradient final focusing permanent magnet quadrupoles, we expect to be able to create the conditions \cite{jlim} for achieving a collapsed-ion equilibrium at the FACET-II interaction point. These experiments should be able to demonstrate propagation of a joint Bennett-type equilibrium.

In these experiments, the plasma needed is within the current state-of-the-art \cite{Barber}. It has a nominal plasma density of $10^{18}$ cm$^{-3}$, in order to provide high phase advance for the ions within the beam. This is also accomplished using a slightly relaxed bunch length of $\sigma_z=40$ $\mu$m. Thus, as seen in the simulations above, we anticipate a Bennett-type equilibrium will develop over nearly the entire beam. This model situation is ideal for studying the physics of ion collapse.

The beam at final focus will have a spot size that is well sub-optical. We will need to utilize both appearance intensity \cite{PlasmaMonitorPRX} and ionization dynamics (as in the E-317 proposal on high field atomic physics at FACET-II, see Ref. \cite{tevpermeter}), In order to resolve these spot sizes (\~ 100 nm) one may use advanced imaging methods \cite{Sukhikh2017}, including coherent imaging approaches \cite{MarinelliCDI}, as well as  other tools developed at SLAC at previous test facilities, \textit{e.g.} laser-wire measurements \cite{Laserwire}. Electron beam-derived forces produce ions having kinetic energy up to the 100 keV range that may be collected to give a signature of ion motion, as well as a  measure of the electron beam size. It is planned to build a short ion transport line to extract the ions away from the electron beam path, and to measure their energies with a compact magnetic spectrometer made from a permanent magnet dipole \cite{PMD2017}. 

The betatron radiation spectra emitted by the electron beam for the parameters given above show unique signatures of gamma-ray production. For the parameters used in simulation, photons are produced well into the MeV range, and the spectrum is broadened by the particular nonlinear focusing of the Bennett-type profile’s fields. The Lenard-Wiechert field-based simulations are now underway at UCLA \cite{SakaiICS2017} to provide detailed theoretical predictions of the spectral shape.  Measurement of the MeV-scale double-differential spectrum will be accomplished  by use of an Compton-pair spectrometer under development at UCLA \cite{Naranjo2020}.

A central goal of the experiment should also verify the predicted emittance growth (near a factor of two) under ion collapse conditions, which should be due predominantly to the phase space filamentation during Bennett-type near-equilibrium formation. This entails measurement of the electron beam downstream of the plasma source. Sensitive emittance measurement techniques are now being developed for the E-310 Trojan Horse Injection experiment \cite{Deng2019}, for deducing emittance growth through beam profile measurements downstream of the plasma exit.

In summary, the physics of critically important phenomena in PWFA have been explored in a theoretical and simulational approach placed in the context of both FACET-II and LC experimental concerns. The issues raised here provide for a rich physics study of ion collapse and beam-ion equilibrium formation. Successful exploration of the physical effects involved should give confidence to the developing view of a PWFA-based future linear collider. 

\section*{Acknowledgments}

This work was performed with support of the US Dept. of Energy, Division of High Energy Physics, under contract no. DE-SC0009914, and the National Science Foundation, under grant no. PHY-1549132, an NSF Science and Technology, the Center for Bright Beams.

This work used computational and storage services associated with the Hoffman2 Shared Cluster provided by UCLA Institute for Digital Research and Education’s Research Technology Group.
 
\hspace{12pt}

\appendix
\section{Derivation of Multiple Scattering Angle Distribution}
\label{sec:multiplescattering}

The derivation of the multiple scattering angular distribution is similar to that found in Ref.~\cite{scott} and Ref.~\cite{bethe}. The major difference is that the Mott-Born differential cross section with a cutoff at angles below a minimum angle $\theta_{\text{min}}$ is used instead of a cross section with atomic screening.

Let $\theta$ be the polar angle and $\beta$ be the azimuthal angle. $\phi_x$ and $\phi_y$ are the projected angles in $x$ and $y$ respectively. Assuming scattering angles are small, these angles are related by

\begin{equation}
\begin{array}{rcr}
     \phi_x &=& \theta \cos(\beta)  \\
     \phi_y &=& \theta \sin(\beta) 
\end{array}.
\end{equation}

Let $F(\theta, \beta)$ be a scattering angle distribution written as a function of the polar and azimuthal angles. The scattering angle distribution as a function of the projected angles is given by $F_p(\phi_x, \phi_y) = F(\theta \cos(\beta), \theta \sin(\beta))$. These distributions are normalized so that

\begin{equation}
\int_0^{2 \pi} \int_0^{\infty} F(\theta, \beta) \theta d\theta d\beta = 1
\label{eq:normcond}
\end{equation}

\noindent and

\begin{equation}
\int_{-\infty}^{\infty} \int_{-\infty}^{\infty} F_p(\phi_x, \phi_y) d\phi_x d\phi_y = 1.
\end{equation}

Note that when the small angle approximation is made, the domain of $\theta$ is extended from $[0, \pi]$ to $[0, \infty)$ and the domains of $\phi_x$ and $\phi_y$ are extended from $[-\pi, \pi]$ to $(-\infty, \infty)$. Now let $F_{p,M}(\phi_x, \phi_y)$ be the distribution of scattering angles after $M$ scatters. The distribution of scattering angles after $M+1$ scatters is given by the convolution

\begin{widetext}

\begin{equation}
F_{p,M+1}(\phi_x, \phi_y) = \int_{-\infty}^{\infty} \int_{-\infty}^{\infty} F_{p,M}(\phi'_x, \phi'_y) F_{p,1}(\phi_x - \phi'_x, \phi_y - \phi'_y) d\phi_x d\phi_y.
\end{equation} 

\noindent In Fourier space, this equation is

\begin{equation}
F_{p,M+1}(\xi_x, \xi_y) = F_{p,M}(\xi_x, \xi_y) F_{p,1}(\xi_x, \xi_y)
\label{eq:fspacemult}
\end{equation}

\noindent where

\begin{equation}
F_{p,M}(\xi_x, \xi_y) = \int_{-\infty}^{\infty} \int_{-\infty}^{\infty} e^{i(\xi_x \phi_x + \xi_y \phi_y)} F_{p,M}(\phi_x, \phi_y) d\phi_x d\phi_y
\label{eq:ftransform}
\end{equation}

\noindent and

\begin{equation}
F_{p,M}(\phi_x, \phi_y) = \frac{1}{(2\pi)^2} \int_{-\infty}^{\infty} \int_{-\infty}^{\infty} e^{-i(\xi_x \phi_x + \xi_y \phi_y)} \tilde{F}_{p,M}(\xi_x, \xi_y) d\xi_x d\xi_y.
\end{equation}

\noindent From Eq.~(\ref{eq:fspacemult}) it can be seen that

\begin{equation}
F_{p,M}(\xi_x, \xi_y) = F_{p,1}(\xi_x, \xi_y)^M.
\label{eq:fspacepower}
\end{equation}

\noindent We now define $\tilde{F}_{M}(\xi,\alpha)$ as 

\begin{equation}
\tilde{F}_{M}(\xi,\alpha) \equiv \tilde{F}_{M}(\xi \cos(\alpha), \xi \sin(\alpha)).
\label{eq:fspacedeffromprojected}
\end{equation}

\noindent Substituting in Eq.~(\ref{eq:ftransform}),

\begin{equation}
\begin{split}
\tilde{F}_M(\xi,\alpha) &= \int_0^{2 \pi} \int_0^{\infty} e^{i\xi\theta(\cos(\alpha) \cos(\beta) + \sin(\alpha) \sin(\beta)} F_{p,M}(\theta \cos(\beta), \theta \sin(\beta) \theta d\theta d\beta \\
&= \int_0^{2 \pi} \int_0^{\infty} e^{i\xi\theta\cos(\beta - \alpha)} F_{M}(\theta, \beta) \theta d\theta d\beta \\
&= \int_0^{\infty} F_M(\theta) \left[\int_0^{2 \pi} e^{i \xi \theta \cos(\beta - \alpha)} d\beta \right] \theta d\theta \\
&= 2 \pi \int_0^{\infty} F_M(\theta) J_0(\xi \theta) \theta d\theta
\end{split}
\label{eq:hankel}
\end{equation}

\end{widetext}

\noindent where we have exploited the fact that the angular distribution is cylindrically symmetric and thus $F_{M}(\theta, \beta) = F_{M}(\theta)$. Note that from Eqs.~(\ref{eq:hankel}), $\tilde{F}_M(\xi,\alpha) = \tilde{F}_M(\xi)$ is independent of $\alpha$. Eq.~(\ref{eq:hankel}) is a Hankel transform of order zero, and the inverse transform is

\begin{equation}
F_M(\theta) = \frac{1}{2\pi} \int_0^{\infty} \tilde{F}_M(\xi) J_0(\theta \xi) \xi d\xi.
\end{equation}

The projected multiple scattering angle distribution can be written in terms of $F_{p,M}(\phi_x, \phi_y)$, the projected scattering angle distribution for $M$ scatters, and $p_M$, the probability of a particle scattering $M$ times, as

\begin{equation} \label{eq:msadsum}
F_p(\phi_x, \phi_y) = \sum_{M=0}^{\infty} F_{p,M}(\phi_x, \phi_y) p_M.
\end{equation}

\noindent To find $p_M$, we imagine a particle which travels an infinitesimal distance $dz$ in a material with number density $n$ of scattering targets each with total cross section $\sigma$. The probability that the particle scatters is $ \sigma n dz$. The probability that it does not scatter is $1 - \sigma n dz$. If we now let the particle travel a finite distance $z$ in the, we can split $z$ into $N$ infinitesimal distances of length $dz = z / N$. $p_M$ is is given by

\begin{equation}
\begin{split}
p_M &= \lim_{N \to \infty} \binom{N}{M} \left( \frac{\sigma n z}{N} \right)^M \left( 1 - \frac{\sigma n z}{N} \right)^{N - M} \\ &= \frac{(\sigma n z)^M}{M!} e^{-\sigma n z} = \frac{\Omega_0^M}{M!} e^{-\Omega_0}
\end{split}
\label{eq:probm}
\end{equation}

\noindent where we have introduced the dimensionless parameter $\Omega_0 \equiv \sigma n z$. Now we Fourier transform both sides of Eq.~(\ref{eq:msadsum}) to give

\begin{equation}
\tilde{F}_p(\xi_x, \xi_y) = \sum_{M=0}^{\infty} \tilde{F}_{p,M}(\xi_x, \xi_y) p_M.
\end{equation}

\noindent Substituting in Eqs.~(\ref{eq:fspacepower}) and (\ref{eq:probm}), we obtain

\begin{equation}
\begin{split}
\tilde{F}_p(\xi_x, \xi_y) &= \sum_{M=0}^{\infty} \tilde{F}_{p,1}(\xi_x, \xi_y)^M \frac{\Omega_0^M}{M!} e^{-\Omega_0} \\ &= e^{(\tilde{F}_{p,1}(\xi_x, \xi_y) - 1) \Omega_0}.
\end{split}
\end{equation}

\noindent Now using Eq.~(\ref{eq:fspacedeffromprojected}) we have

\begin{equation}
\begin{split}
\tilde{F}(\xi) &= \tilde{F}(\xi, \alpha) \\ &= \tilde{F_p}(\xi \cos(\alpha), \xi \sin(\alpha) \\ &= e^{(\tilde{F}_{p,1}(\xi \cos(\alpha), \xi \sin(\alpha)) - 1) \Omega_0} \\ &= e^{(\tilde{F}_1(\xi, \alpha) - 1) \Omega_0} \\ &= e^{(\tilde{F}_1(\xi) - 1) \Omega_0}.
\end{split}
\label{eq:ftildeftilde1}
\end{equation}

The angular distribution after a single scatter $F_1(\theta)$ is proportional to the differential cross section. The constant of proportionality is determined by the normalization condition Eq.~(\ref{eq:normcond}). $F_1(\theta)$ is given by

\begin{equation}
F_1(\theta) = \frac{1}{\sigma} \frac{d\sigma}{d\Omega}(\theta)
\end{equation}

\noindent where it is evident from the definition of total cross section that $F_1(\theta)$ obeys the normalization condition Eq.~(\ref{eq:normcond}). In summary, the multiple scattering angular distribution is given by

\begin{equation}
F(\theta) = \frac{1}{2\pi} \int_0^{\infty} e^{(\tilde{F}_1(\xi) - 1) \Omega_0} J_0(\theta \xi) \xi d\xi,
\label{eq:msadhankel}
\end{equation}

\noindent where

\begin{equation}
\tilde{F}_1(\xi) = 2\pi \int_0^{\infty} \frac{1}{\sigma} \frac{d\sigma}{d\Omega}(\theta) J_0(\xi \theta) \theta d\theta.
\label{eq:tildef1}
\end{equation}

We now evaluate these equations using the Mott-Born differential cross section with a cutoff at angles below a minimum angle $\theta_{\text{min}}$. The differential cross section is

\begin{equation}
\frac{d\sigma}{d\Omega}(\theta) = 
\begin{cases}
    \frac{Z^2 r_e^2 (1 - \beta_0^2\sin^2(\frac{\theta}{2}))}{4 \gamma_0^2 \beta_0^4 \sin^4(\frac{\theta}{2})} & \theta > \theta_{\text{min}}\\
    0 & \theta < \theta_{\text{min}},
\end{cases}
\end{equation}

\noindent where $Z$ is the atomic number of the ions and $r_e$ is the classical electron radius. The total cross section is

\begin{equation}
\begin{split}
\sigma &= \int_0^{2\pi} \int_{0}^{\pi} \frac{d\sigma}{d\Omega}(\theta) \sin(\theta) d\theta d\beta .\\
&= \frac{\pi r_e^2 Z^2}{\beta_0^4 \gamma_0^2} \left[\cot \left( \frac{\theta_{\text{min}}}{2} \right)^2 + \ln \left( \sin \left ( \frac{\theta_{\text{min}}}{2} \right) \right) \right], \\
& \simeq \frac{4\pi r_e^2 Z^2}{\beta_0^4 \gamma_0^2 \theta_{\text{min}}^2}, 
\end{split}
\label{eq:totalcrosssection}
\end{equation}

\noindent where we have assumed $\theta_{\text{min}}\ll 1$. We write the total cross section in terms of the cross-sectional radius $r_\sigma \equiv \sqrt{\sigma / \pi}$. In terms of $r_\sigma$, the minimum angle is

\begin{equation}
\theta_{\text{min}} = \frac{2 Z r_e}{\gamma_0 \beta_0^2 r_\sigma}.
\end{equation}

\noindent The series expansion of $\frac{d\sigma}{d\Omega}(\theta)$ about $\theta = 0$ is

\begin{equation}
\frac{d\sigma}{d\Omega}(\theta) \simeq 
r_\sigma^2 \theta_{\text{min}}^2
\begin{cases}
    \frac{1}{\theta^4} + \frac{2 - 3\beta_0^2}{12 \theta^2} + \text{O}(\theta^0)& \theta > \theta_{\text{min}}\\
    0 & \theta < \theta_{\text{min}}
\end{cases}.
\end{equation}

\noindent From Eqs.~(\ref{eq:tildef1}) and (\ref{eq:totalcrosssection}),

\begin{equation}
\begin{split}
\tilde{F}_{1}(\xi) &= \frac{2\pi}{\sigma} \int_0^{\infty} \frac{d\sigma}{d\Omega}(\theta) J_0(\xi \theta) \theta d\theta \\
&= 2\theta_{\text{min}}^2 \int_{\theta_{\text{min}}}^{\infty} \left(\frac{1}{\theta^3} + \frac{2 - 3\beta_0^2}{12 \theta} \right) J_0(\xi \theta) d\theta \\
&= 2\theta_{\text{min}}^2 \xi^2 \int_{\xi \theta_{\text{min}}}^{\infty} \frac{J_0(u)}{u^3} du \\
&+ \frac{(2-3\beta_0^2)\theta_{\text{min}}^2 \xi}{6} \int_{\xi \theta_{\text{min}}}^{\infty} \frac{J_0(u)}{u} du \\
&= 2x^2 I_1(x) + \frac{(2-3\beta_0^2)\theta_{\text{min}}}{6} x I_2(x)
\end{split}
\label{eq:f1}
\end{equation}

\noindent where $x \equiv \theta_{\text{min}} \xi$ and the integrals $I_1(x)$ and $I_2(x)$ are defined as

\begin{equation}
I_1(x) \equiv \int_x^{\infty}\frac{J_0(u)}{u^3} du,
\end{equation}

\noindent and 

\begin{equation}
I_2(x) \equiv \int_x^{\infty}\frac{J_0(u)}{u} du.
\label{eq:i2}
\end{equation}

\noindent Integrating by parts twice and using the relations $(J_0(x))' = -J_1(x)$ and $(x J_1(x))'= x J_0(x)$, $I_1(x)$ can be written

\begin{equation}
I_1(x) = \frac{J_0(x)}{2x^2} - \frac{J_1(x)}{4x} - \frac{1}{4}I_2(x).
\end{equation}

\noindent Eq.~(\ref{eq:f1}) now becomes

\begin{equation}
\begin{split}
\tilde{F}_1(\xi) &= J_0(x) - \frac{x}{2} J_1(x) \\ &+ \left(\frac{(2-3\beta_0^2)\theta_{\text{min}} x}{6} - \frac{x^2}{2} \right) I_2(x).
\end{split}
\label{eq:f1blah}
\end{equation}

\noindent Eq.~(\ref{eq:i2}) can be simplified to

\begin{equation}
\begin{split}
I_2(x) &= \frac{2}{\pi} \int_x^{\infty} \int_0^{\frac{\pi}{2}} \frac{\cos(u \cos(\phi))}{u} d\phi du \\ &= \frac{2}{\pi} \int_0^{\frac{\pi}{2}} \int_{x \cos(\phi)}^{\infty} \frac{\cos(v)}{v} dv d\phi \\ &= -\frac{2}{\pi} \int_0^{\frac{\pi}{2}} \text{Ci}(x \cos(\phi)) d\phi \\
&= -\left[\ln \left(\frac{x e^{\gamma_E}}{2} \right) + \sum_{n=1}^{\infty} \frac{(-1)^n x^{2n}}{2 n 4^n (n!)^2} \right]
\end{split}
\end{equation}

\noindent where $\gamma_E = 0.57721...$ is the Euler-Mascheroni constant and where we have used

\begin{equation}
J_0(x) = \frac{2}{\pi} \int_0^{\frac{\pi}{2}} \cos(x \cos(\phi)) d\phi,
\end{equation}

\begin{equation}
\text{Ci}(x) = -\int_x^{\infty} \frac{\cos(u)}{u} du,
\end{equation}

\noindent and

\begin{equation}
\text{Ci}(x) = \gamma + \ln(x) + \sum_{n=1}^{\infty}\frac{(-x^2)^n}{2n (2n)!}.
\end{equation}

From Ref.~\cite{scott} and Ref.~\cite{bethe} the important contribution to the integral Eq.~(\ref{eq:msadhankel}) occurs when $x \lesssim 1$. Substituting the series expansions of $J_0(x)$, $J_1(x)$, and $I_2(x)$ in Eq.~(\ref{eq:f1blah}) and ignoring the $\text{O}(x^2)$ and larger terms, and using the fact that $\theta_{\text{min}}\ll 1$, we obtain

\begin{equation}
\tilde{F}_{1}(\xi) = 1 + \frac{x^2}{2} \ln \left(\frac{x e^{\gamma_E}}{2} \right) + \text{O}(x^2).
\end{equation}

Let $\eta \equiv \sqrt{\Omega_0 B} x$ where $B$ is a constant defined later.

\begin{equation}
\begin{split}
\tilde{F}_{1}(\xi) - 1 &= \frac{\eta^2}{4 \Omega_0 B} \ln \left( \frac{\eta^2  e^{2 \gamma_E}}{4 \Omega_0 B e^2} \right) \\
&= \frac{\eta^2}{4 \Omega_0 B} \left( \ln \left( \frac{e^{2 \gamma_E}}{ \Omega_0 B e^2} \right)  + \ln \left( \frac{\eta^2}{ 4} \right) \right)
\end{split}
\label{eq:etacoordtransform}
\end{equation}

\noindent If we choose $B$ such that

\begin{equation}
\frac{1}{B} \ln \left( \frac{e^{2\gamma_E}}{\Omega_0 B e^2} \right) = -1
\label{eq:bdef}
\end{equation}

\noindent or equivalently

\begin{equation}
B = -W_{-1}\left( \frac{-e^{-2(1 - \gamma_E)}}{\Omega_0} \right)
\end{equation}

\noindent where $W_{-1}$ is the lower branch of the Lambert W function, Then by combining Eqs.~(\ref{eq:ftildeftilde1}), (\ref{eq:etacoordtransform}), and (\ref{eq:bdef}), we have

\begin{equation}
\begin{split}
\tilde{F}(\xi) &= e^{- \frac{\eta^2}{4}  + \frac{\eta^2}{4 B} \ln \left( \frac{\eta^2}{ 4} \right)} \\
&\simeq e^{-\frac{\eta^2}{4}} \sum_{n=1}^{\infty} \frac{1}{n!} \left(\frac{\eta^2}{4 B} \ln \left( \frac{\eta^2}{4} \right)\right)^{n} 
\end{split}
\end{equation}

\noindent Plugging this into Eq.~(\ref{eq:msadhankel}) and defining  $\vartheta \equiv \theta / (\theta_{\text{min}} \sqrt{\Omega_0 B})$, the probability density function is given by 

\begin{equation} 
f(\vartheta) = 2\vartheta e^{-\vartheta^2} + \sum_{n=1}^{\infty}\frac{f^{(n)}(\vartheta)}{B^n};
\end{equation}

\noindent where $2\pi f(\vartheta) \vartheta d\vartheta = 2\pi F(\theta) \theta d\theta$ and

\begin{equation} 
f^{(n)}(\vartheta) = \frac{\vartheta }{n!}\int_0^{\infty} e^{\frac{-\eta^2}{4}}\left( \frac{\eta^2}{4} \ln \left( \frac{\eta^2}{4} \right) \right)^n J_0(\eta \vartheta) \eta d\eta.
\end{equation}

\section{Equilibrium Simulation Code Tracking Validation} \label{sec:trackingvalidation}

\begin{figure}[htb]
\centering
\includegraphics[width=\columnwidth]{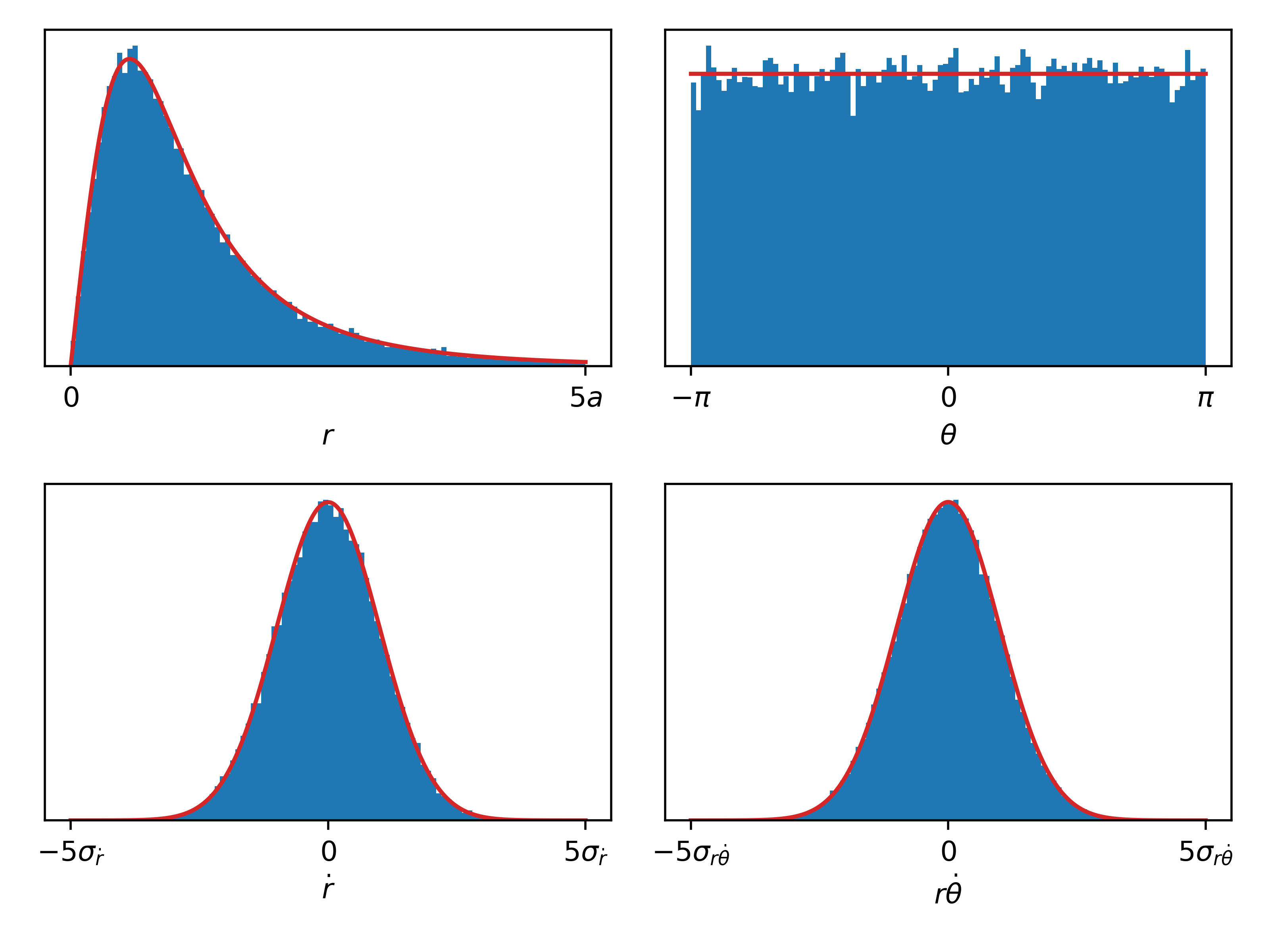}
\caption{Histogram of tracked particles at the end of the plasma (blue) compared to the theoretical particle distribution (red). This simulation was run with the LC parameters shown in Table \ref{tab:eqsimlcparams} except only 50,000 particles were tracked. Scattering was not included in the simulation.}
\label{fig:tvalidation}
\end{figure}

By definition, if a beam is sampled from the equilibrium distribution function and tracked according to the equations of motion, it will remain distributed according to the equilibrium distribution function. Thus the tracking code can be validated by comparing the distribution of beam after it has been tracked to the theoretical distribution function. Including acceleration in the tracking allows the gamma scaling of the equilibrium distribution parameters to be checked.

In Fig.~\ref{fig:tvalidation}, we compare the equilibrium distribution function to the simulated particle distribution at the end of the plasma in the linear collider simulation described in Sec.~(\ref{sec:tcresults}). Because the simulated distribution matches the theoretical distribution in the figure, we can conclude that both the tracking algorithm correctly tracks particles through accelerating fields.

\section{Equilibrium Simulation Code Scattering Validation} \label{sec:scatteringvalidation}

\begin{figure}[htb]
\centering
\includegraphics[width=\columnwidth]{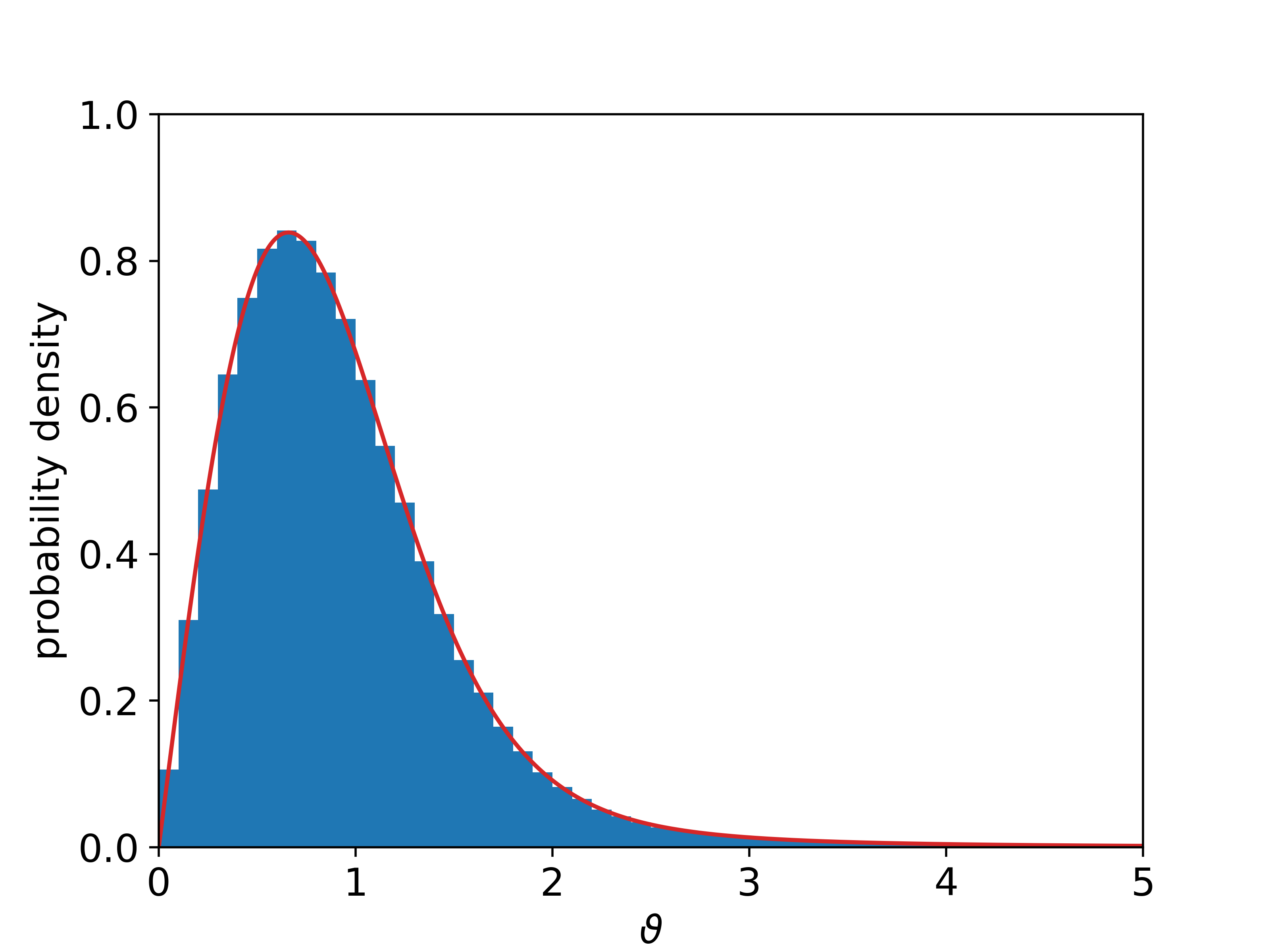}
\caption{Histogram of 1,000,000 values of $\vartheta$ sampled from Eq.~(\ref{eq:distsum}) with $B = 10$ (blue) compared to the probability density function Eq.~(\ref{eq:distsum}) (red).}
\label{fig:samplehist}
\end{figure}

\begin{figure}[htb]
\centering
\includegraphics[width=\columnwidth]{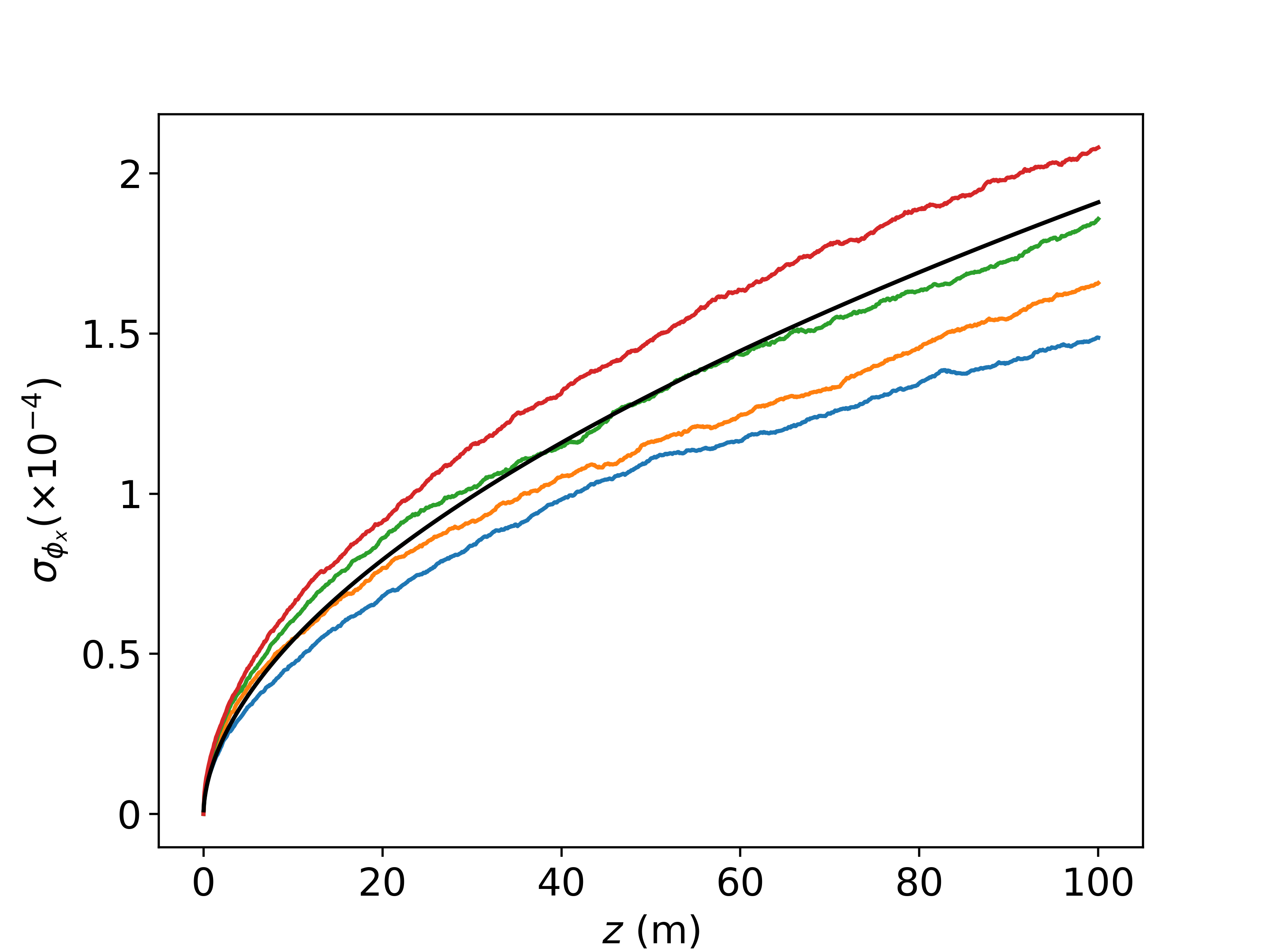}
\caption{Angular spread of electrons tracked through hydrogen gas for differing values of $r_{\sigma}$: $10^{-10} \si{m}$ (blue), $10^{-9} \si{m}$ (orange), $10^{-8} \si{m}$ (green), and $10^{-7} \si{m}$ (red). These simulation results are compared to the theoretical value of $\sigma_{\phi}$ from Eq.~(\ref{eq:molieretheory}) shown in black.}
\label{fig:validationrsigmascan}
\end{figure}

\begin{figure}[htb]
\centering
\includegraphics[width=\columnwidth]{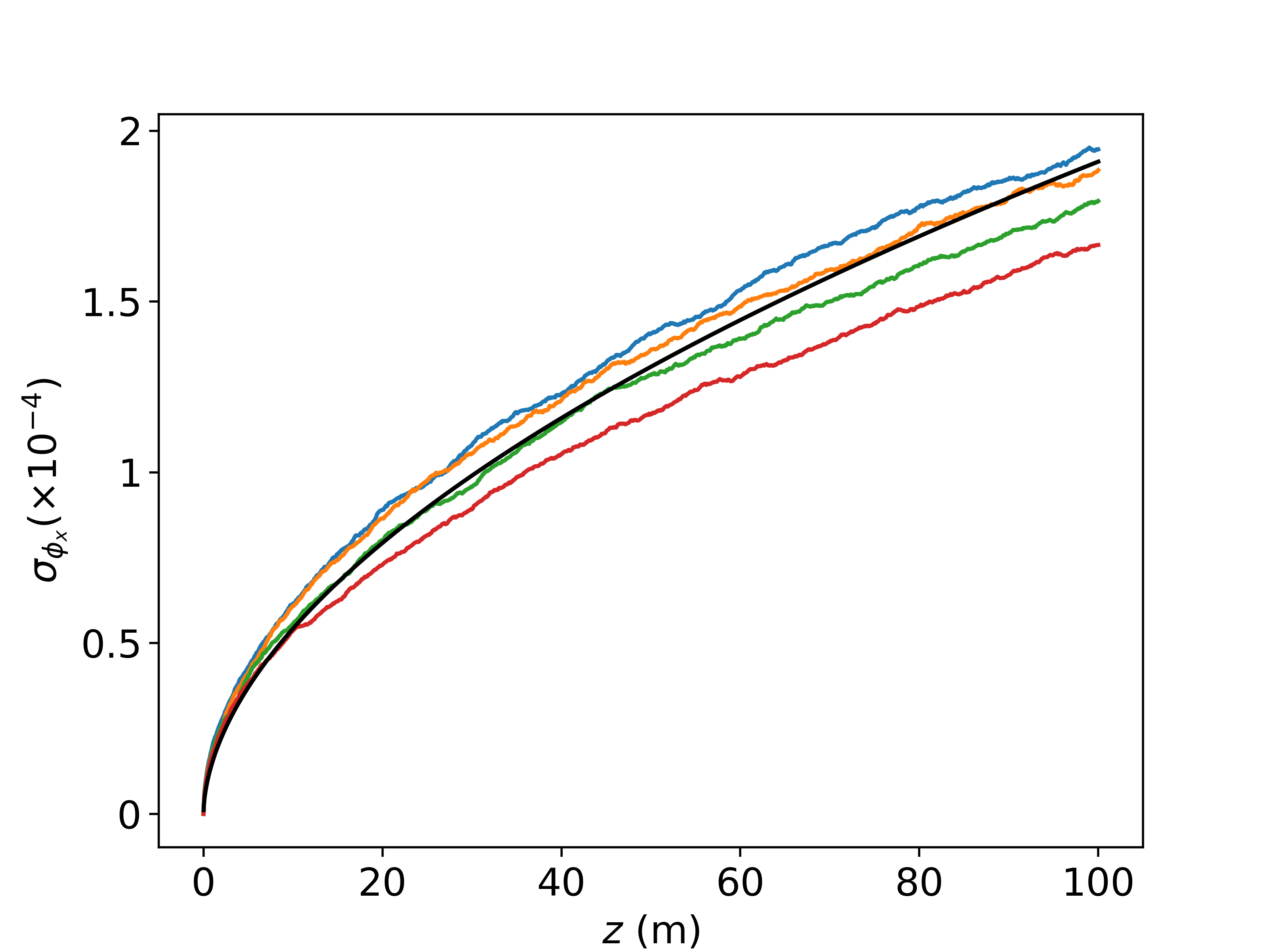}
\caption{Angular spread of electrons tracked through hydrogen gas for differing number of steps: $10^3$ (blue), $10^4$ (orange), $10^5$ (green), and $10^6$ (red). These simulation results are compared to the theoretical value of $\sigma_{\phi}$ from Eq.~(\ref{eq:molieretheory}) shown in black.}
\label{fig:validationstepscan}
\end{figure}

The first step taken to validate the scattering algorithm was to check that the code correctly samples the scattering angles from the distribution given by Eq.~(\ref{eq:distsum}). This is necessary due to the complexity of the algorithm which is described in Sec.~(\ref{sec:trackingcoulombscattering}). First, to ensure the values of Eq.~(\ref{eq:distsum}) computed by the code were correct, they were compared to values computed using Wolfram Mathematica. These were found to be in agreement for various values of $\vartheta$. The next step was to check that the sampled angles are distributed according to the probability density function Eq.~(\ref{eq:distsum}). This was done by arbitrarily choosing $B = 10$ and sampling 1,000,000 values of $\vartheta$ and plotting a histogram of these values against the probability density function. This is shown in  Fig.~\ref{fig:samplehist} and it is clear that the sampled values are distributed according to the correct probability density function.

For the second test of the scattering algorithm, the code was modified to track electrons initialized on axis with zero transverse momentum through a uniform hydrogen gas of density $n_0$ with no forces due to electromagnetic fields. To zeroth order in this situation, Moli\`ere's theory predicts that the projected angles $\phi_x$ and $\phi_y$ are normally distributed with standard deviation given by \cite{revparticlephysics}:

\begin{equation} \label{eq:molieretheory}
\sigma_\phi = \frac{26.6}{\gamma} \sqrt{\frac{m_p n_0 z}{X_0}} \left( 1 + 0.038 \ln \left(\frac{m_p n_0 z}{X_0} \right) \right)
\end{equation}

where $X_0$ is the radiation length which is $630.4 \si{\frac{m^2}{kg}}$ for hydrogen. Besides those such as the integration tolerance used in the calculation of $f(\vartheta)$, there are two unphysical parameters in the scattering algorithm: the cross sectional radius $r_\sigma$ and the number of steps. First the modified code was run with 1000 particles at 10 \si{GeV} tracked in 10000 steps through a 100 \si{m} plasma of density $10^{20} \si{cm^{-3}} $ with the cross sectional radius scanned from $10^{-10} \si{m}$ to $10^{-7} \si{m}$. The simulation results are plotted against theory in Fig.~\ref{fig:validationrsigmascan}. From these results we can see that the scattering algorithm is more or less in agreement with the theoretical estimate in Eq.~(\ref{eq:molieretheory}) for $r_\sigma \approx 10 \si{nm}$. Next the same simulation was run except this time the cross sectional radius was fixed at $10 \si{nm}$ and the number of steps was scanned from 10 to 1,000,000. The simulation results are plotted against theory in Fig.~\ref{fig:validationstepscan}. From this plot we can see that the growth in $\sigma_\theta$ was only slightly affected by changing the number of steps by four orders of magnitude. The number of steps in the simulations performed for the FACET-II and LC cases are approximately 3,000,000. From the results shown in Fig.~\ref{fig:validationstepscan}, we can say the scattering algorithm is a mild underestimate of the scattering predicted by Eq.~(\ref{eq:molieretheory}). However not too much faith should be vested in Eq.~(\ref{eq:molieretheory}) as it is a zeroth order approximation.

\bibliography{refs}

\end{document}